\documentclass[PRM,twocolumn,groupedaddress,nofootinbib,superscriptaddress,longbibliography]{revtex4-2}
\usepackage[version=3]{mhchem}
\usepackage{hyperref}
\usepackage{graphicx}
\usepackage{amsmath,amsfonts}
\usepackage{dcolumn}
\usepackage{bm}
\usepackage{color}
\usepackage{multirow}
\usepackage{physics}
\usepackage{subfigure}
\usepackage[utf8]{inputenc}
\usepackage{titlesec}
\usepackage{float}
\usepackage[margin=0.6in]{geometry}
\usepackage[dvipsnames]{xcolor}

\newcommand{\Tb}{Tb$_2$IrAl$_4$Ge$_2$}
\newcommand{\Er}{Er$_2$IrAl$_4$Ge$_2$}
\newcommand{\Ln}{$Ln_2$IrAl$_4$Ge$_2$}

\begin{document}


\clearpage

\title{Complex magnetic phase diagrams in Tb$_2$IrAl$_4$Ge$_2$ and Er$_2$IrAl$_4$Ge$_2$} 

\author{Karolina Gornicka}
\email{gornickaka@ornl.gov}
\affiliation{Materials Science and Technology Division, Oak Ridge National Laboratory, Oak Ridge, Tennessee 37831, United States}
\affiliation{Faculty of Applied Physics and Mathematics and Advanced Materials Centre, Gdansk University of Technology, ul. Narutowicza 11/12, 80-233 Gdańsk, Poland}

\author{Matthew S. Cook}
\affiliation{Materials Science and Technology Division, Oak Ridge National Laboratory, Oak Ridge, Tennessee 37831, United States}

\author{Brenden R. Ortiz}
\affiliation{Materials Science and Technology Division, Oak Ridge National Laboratory, Oak Ridge, Tennessee 37831, United States}

\author{Andrew D. Christianson}
\affiliation{Materials Science and Technology Division, Oak Ridge National Laboratory, Oak Ridge, Tennessee 37831, United States}

\author{Andrew F. May}
\affiliation{Materials Science and Technology Division, Oak Ridge National Laboratory, Oak Ridge, Tennessee 37831, United States}

\date{\today}

\begin{abstract}
We report the synthesis and comprehensive investigation of single-crystalline \Tb~ and \Er, the first Ir-based members of the $Ln_2M$Al$_4$Ge$_2$ family. Both compounds crystallize in the tetragonal \ce{Tb2NiAl4Ge2}-type structure (space group $I4/mmm$) and exhibit easy-axis magnetic anisotropy with the crystallographic $c$ axis as the preferred magnetization direction. \Tb~ undergoes antiferromagnetic ordering below $T_N = 25$ K and develops a sequence of field-induced metamagnetic transitions accompanied by pronounced magnetotransport anomalies, including a nonlinear Hall response. In contrast, \Er~ exhibits three successive zero-field magnetic transitions and a substantially richer low-temperature magnetic phase diagram comprising multiple competing field-induced phases confined to relatively modest magnetic fields. The differences between the Tb- and Er-based compounds demonstrate the wide range of magnetic behavior that can be realized within the \ce{Tb2NiAl4Ge2}-type structure. More broadly, the successful realization of Ir-based $Ln_2M$Al$_4$Ge$_2$ compounds considerably expands the accessible chemical space of this family and opens a route toward the exploration of related $4d$- and $5d$-based aluminogermanides.
\end{abstract}


\maketitle

\section{Introduction}

Complex magnetic phase diagrams are a hallmark of many rare-earth intermetallic compounds. Their magnetic behavior is often governed by a delicate balance between long-range exchange interactions and single-ion anisotropy. In particular, the interplay between Ruderman-Kittel-Kasuya-Yosida (RKKY) exchange interactions, which couple localized $4f$ moments via conduction electrons, crystal-electric-field (CEF) effects, and the Zeeman energy can stabilize a wide variety of magnetic ground states, ranging from simple magnetic order to complex sequences of field-induced phases \cite{ruderman1954indirect, van1962note, kasuya1956theory}. As a result, rare-earth intermetallics continue to serve as an important platform for investigating competing magnetic interactions and field-driven phenomena \cite{xie2021complex,wu2024transition,wu2026possible,gignoux1993competition,feng2026complex,khan2019steplike,ram2023multiple,sharma2022complex,gamza2019complex,yang2025two}.

Within this broad class of materials, multicomponent intermetallic compounds are especially attractive because their increased chemical complexity provides additional structural flexibility and enables the formation of diverse atomic arrangements that are often inaccessible in simpler binary or ternary systems. Consequently, quaternary intermetallics considerably expand the available structural and chemical space for tuning electronic and magnetic properties, making them valuable candidates for exploring structure-property relationships. Among such systems, rare-earth aluminogermanides have recently emerged as a particularly fertile platform for complex magnetism \cite{feng2022magnetic,zhang2023multiple,wu2005reaual4ge2,rezende2025metallic,feng2024complex,leahy2022field,feng2023revealing,cong2023magnetic,wu2025collinear,zhang2017electronic}. A prominent example is the $LnM$Al$_4$Ge$_2$ family, where $Ln$ denotes a lanthanide element and $M$ is a transition metal. These compounds crystallize in the rhombohedral space group $R\bar{3}m$ \cite{wu2005reaual4ge2} and feature well-separated triangular nets of magnetic rare-earth ions. The combination of competing magnetic interactions, geometric frustration, and magnetic anisotropy gives rise to rich magnetic phase diagrams and a variety of field-induced phenomena. Representative members include the Ni-, Pt-, Pd-, and Au-based compounds, which exhibit multiple magnetic phase transitions, unconventional magnetotransport responses, and, in some cases, have been proposed as candidates for hosting nontrivial magnetic textures \cite{shin2020magnetic,leahy2022field,zhang2017electronic,cong2023magnetic,gao2024high,feng2022magnetic,wu2025collinear}.

A less-explored branch of quaternary aluminogermanides is the $Ln_2M$Al$_4$Ge$_2$ family, which crystallizes in the tetragonal \ce{Tb2NiAl4Ge2}-type structure (space group $I4/mmm$) \cite{sieve2002quaternary,demchenko2008quaternary,semuso2015tokaychuk}. In contrast to the rhombohedral $R\bar{3}m$ compounds, these materials contain alternating \ce{$M$Al4Ge2} slabs and rare-earth bilayers, resulting in a distinctly three-dimensional rare-earth sublattice. To date, this structure type has been reported primarily for compounds containing the $3d$ transition metals Fe, Co, and Ni. Structural investigations have established extensive isostructural series across the rare-earth elements ($Ln$ = Y, Sm, Gd–Lu), while systematic investigations of their physical properties remain relatively limited \cite{he2015crystal,demchenko2008quaternary,pukas2024quaternary,huang2019physical,sieve2002quaternary}. The magnetic studies currently available point to predominantly antiferromagnetic behavior throughout the family, with long-range magnetic order reported for compounds including \ce{Gd2CoAl4Ge2} \cite{he2015crystal,huang2019physical}, \ce{Gd2FeAl4Ge2} \cite{he2015crystal}, and \ce{Tb2NiAl4Ge2} \cite{sieve2002quaternary}.  Moreover, the majority of reported compounds have been prepared and characterized in polycrystalline form, which restricts access to anisotropic properties and field-dependent phenomena, leaving the broader magnetic landscape of the $Ln_2M$Al$_4$Ge$_2$ family largely unexplored.

In this work, we report the synthesis, crystal structure, and comprehensive investigation of the magnetic, thermodynamic, and transport properties of single-crystalline \Tb~ and \Er~, the first members of the $Ln_2M$Al$_4$Ge$_2$ family containing the $5d$ transition metal Ir. The successful growth of these single crystals demonstrates that this structure type remains stable beyond the previously reported $3d$ transition-metal compounds, opening a pathway toward the exploration of additional $4d$- and $5d$-based members of the $Ln_2M$Al$_4$Ge$_2$ family. The replacement of a $3d$ element with the heavier $5d$ metal introduces a substantially different electronic environment and raises the possibility of enhanced spin-orbit coupling effects. We show that both compounds undergo antiferromagnetic ordering and exhibit remarkably rich field-induced phase diagrams characterized by multiple metamagnetic transitions, and pronounced signatures in the magnetotransport response. These results establish the Ir-based members of the $Ln_2M$Al$_4$Ge$_2$ family as promising candidates for future investigations of field-induced magnetism and motivate further exploration of related $4d$- and $5d$-based compounds.
 
\section{Experimental details}
\begin{figure}[t]
\centering
\includegraphics[width=1\columnwidth]{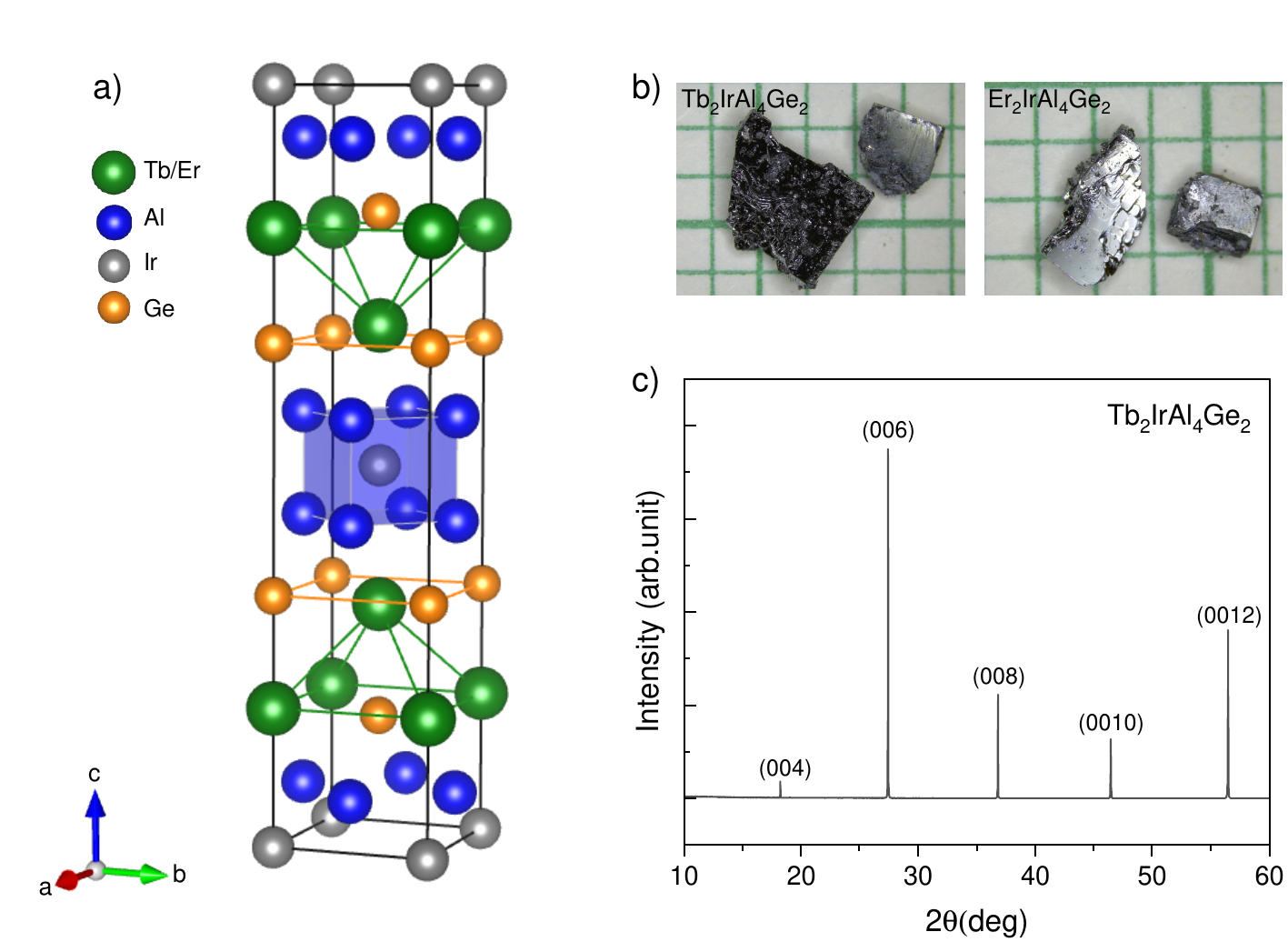}
\caption{(a) Crystal structure of \Ln~ highlighting the arrangement of $Ln$ (green), Al (blue), Ir (grey), and Ge (orange) atoms. (b) Photographs of \Er~ and \Tb~ single crystals, showing their plate-like morphology. (c) X-ray diffraction pattern measured on a flat facet of a \Tb~ crystal, indicating that the crystal surface is parallel to the $ab$ plane, with the $c$ axis perpendicular to the surface.}
\label{Fig.1}
\end{figure}

Single crystals of \Tb~ and \Er~ were grown using an Al-Ge self-flux method. The choice of flux composition was guided by the low-temperature eutectic in the Al–Ge phase diagram, which provides a suitable liquid medium for crystal growth. High-purity elements Tb/Er, Ir, Al, and Ge ($\geq99.99\%$) were weighed in an initial molar ratio of 1:1:10:5 and loaded into 2 mL Canfield crucibles fitted with a catch crucible and a porous frit \cite{canfield2016use}. The crucible sets were sealed in evacuated quartz ampoules after being purged with argon and placed in a box furnace. The temperature profile was adapted, with minor modifications, from the crystal-growth procedure reported by S. \textcolor{BrickRed}{Latturner} et al. for quaternary intermetallics grown from molten aluminum \cite{latturner2002quaternary}. The samples were heated to 1000$^\circ$C at a rate of 100$^\circ$C/h, followed by a dwell of 12 h to ensure homogenization. The melts were cooled to 860$^\circ$C at 7$^\circ$C/h and held for 48 h to improve crystal quality. The temperature was further reduced to 700$^\circ$C at 12$^\circ$C/h, at which point the residual liquid flux was removed by centrifugation. This process yielded plate-like single crystals (see Fig.\ref{Fig.1}b) with typical lateral dimensions of 1–3 mm, which were found to be stable in air and moisture.

Single crystal X-ray diffraction (SCXRD) data were collected on a Bruker D8 Advance Quest diffractometer with a graphite monochromator using Mo K$\alpha$ ($\lambda$=0.71073 \AA). Small, regularly shaped crystals were mounted on Kapton loops using a small amount of GE-varnish. The structural solutions were obtained and refined using the full-matrix least-squares method with the Bruker APEX4 software package. The crystallographic orientation of the as-grown crystals was examined by X-ray diffraction measurements performed on flat crystal facets using a Malvern Panalytical Empyrean X-ray Diffractometer with monochromated Cu K$\alpha_1$ radiation. Elemental analysis was carried out on as-grown crystals using a Hitachi-TM4000 Plus microscope equipped with Oxford’s energy dispersive spectroscopy (EDS) system.

Magnetization measurements were performed using a 7 T Quantum Design MPMS3 SQUID magnetometer, operated in vibrating-sample magnetometry (VSM) mode, in the temperature range of 1.9–300 K. To probe magnetic anisotropy, measurements were carried out on oriented single crystals with the magnetic field applied either parallel to the crystallographic $c$ axis ($H \parallel c$) or within the $ab$ plane ($H \perp c$). Magnetization as a function of temperature, $M(T)$, was measured under both zero-field-cooled (ZFC) warming and field-cooled (FC) cooling protocols for fields up to 1000 Oe. Since no measurable difference between the ZFC and FC data was observed, magnetization at higher fields was collected only upon cooling in the applied field. Field-dependent magnetization, $M(H)$, was measured at various temperatures. For \Er~, low-temperature measurements down to 0.4 K were performed using the QD iHe-3 $^3$He insert for the MPMS3. For the Tb-based compound, additional high-field magnetization measurements up to $12~\mathrm{T}$ were performed using a QD DynaCool Physical Property Measurement System (PPMS) with VSM option.

Heat capacity ($C_p$) measurements were carried out by the standard semiadiabatic relaxation method implemented in a QD PPMS. Measurements were carried out using a standard 2\% temperature rise, with additional field-dependent measurements performed for magnetic fields applied parallel to the crystallographic $c$ axis. Due to experimental constraints related to sample stability on the calorimetry platform under high magnetic fields, the applied field in the heat-capacity measurements was limited to 40 kOe and temperatures above 1.9 K. Additional measurements using a large heat-pulse method with a 30\% temperature rise were also performed to examine the possible presence of first-order transitions. For this purpose, the data were analyzed using a single-slope expression, which is appropriate for first-order phase transitions. For the Er-based compound, additional low-temperature heat-capacity measurements extending into the $^{3}$He temperature range were performed using the QD $^{3}$He insert in zero magnetic field. Electrical resistivity ($\rho$), magnetoresistance (MR) and Hall effect measurements were conducted using the ac resistance (ETO) option on a Dynacool. Platinum electrical leads were attached to the samples using silver paint (DuPont 4929N) in a standard four-point collinear configuration. The electrical current was applied within the $ab$ plane, while the magnetic field was applied parallel to the $c$ axis. For Hall-effect measurements, the raw transverse resistance data were antisymmetrized with respect to magnetic field in order to remove contributions arising from longitudinal magnetoresistance. No additional background subtraction or correction procedures were applied.

\begin{table}[t]
\caption{Results from refinements of single-crystal x-ray diffraction data for \Tb~ and \Er~ collected at ambient conditions.}
\label{tab:scxrd_refinement}
\centering
\begin{tabular*}{\columnwidth}{@{\extracolsep{\fill}}lcc}
\hline\hline
Parameter & \Tb~ & \Er~ \\
\hline
Space group & $I4/mmm$ (139) & $I4/mmm$ (139) \\
$a$ (\AA) & 4.205(2) & 4.180(2) \\
$c$ (\AA) & 19.581(4) & 19.443(4) \\
Volume (\AA$^3$) & 346.2(3) & 339.6(3) \\
Density (g\,cm$^{-3}$) & 7.320 & 7.625 \\
$R_1$ (all data) & 0.0374 & 0.0149 \\
$wR_2$ (all data) & 0.0921 & 0.0314 \\
$R_{\rm int}$ & 0.0630 & 0.0422 \\
Goodness of fit & 1.218 & 1.217 \\
Highest peak/deepest hole
& 3.31/$-$2.45
& 1.35/$-$1.48 \\
Data/parameters & 244/14 & 285/14 \\
\hline\hline
\end{tabular*}
\end{table}

\begin{table}[t]
\caption{Atomic coordinates and equivalent isotropic displacement parameters for \Tb.}
\label{tab:Tb_coords}
\centering
\begin{tabular}{lcccc}
\hline\hline
Atom & $x$ & $y$ & $z$ & $U_{\rm eq}$ (\AA$^2$) \\
\hline
Tb & 0 & 0 & 0.18471(4) & 0.0093(3) \\
Ir & $\frac{1}{2}$ & $\frac{1}{2}$ & $\frac{1}{2}$ & 0.0074(3) \\
Ge & 0 & 0 & 0.33579(10) & 0.0100(4) \\
Al & 0 & $\frac{1}{2}$ & 0.43168(18) & 0.0113(7) \\
\hline\hline
\end{tabular}

\caption{Atomic coordinates and equivalent isotropic displacement parameters for \Er.}
\label{tab:Er_coords}
\centering
\begin{tabular}{lcccc}
\hline\hline
Atom & $x$ & $y$ & $z$ & $U_{\rm eq}$ (\AA$^2$) \\
\hline
Er & 0 & 0 & 0.18510(2) & 0.00508(10) \\
Ir & $\frac{1}{2}$ & $\frac{1}{2}$ & $\frac{1}{2}$ & 0.00299(10) \\
Ge & 0 & 0 & 0.33434(3) & 0.00519(14) \\
Al & $\frac{1}{2}$ & 0 & 0.43101(7) & 0.0063(3) \\
\hline\hline
\end{tabular}
\end{table}

\section{Results and discussion}
\subsection{Structural and chemical characterization}
We begin with the structural characterization of the synthesized single crystals. The chemical composition and homogeneity were verified by EDS measurements performed on multiple crystals and at several spots on each sample. The average atomic ratio, $Ln$:Ir:Al:Ge = 22 $\pm$ 2 : 11 $\pm$ 1 : 46 $\pm$ 3 : 21 $\pm$ 2, is consistent with the stoichiometry of \Ln~ within experimental uncertainty, confirming the expected composition and good chemical homogeneity.

The single-crystal X-ray diffraction results show that both \Er~ and \Tb~ crystallize in the tetragonal \ce{Tb2NiAl4Ge2}-type structure (space group $I4/mmm$, No. 139). The obtained lattice parameters, $a = 4.180(2)$ Å and $c = 19.443(4)$ Å for \Er, and $a = 4.205(2)$ Å and $c = 19.581(4)$ Å for the Tb-based compound, are slightly larger than those reported for the Fe-, Ni-, and Co-based analogs \cite{sieve2002quaternary,he2015crystal,demchenko2008quaternary}, consistent with the larger atomic radius of Ir relative to $3d$ transition metals. The crystallographic parameters and refinement details are summarized in Table~\ref{tab:scxrd_refinement}, while the refined atomic coordinates are listed in Table~\ref{tab:Tb_coords} and Table~\ref{tab:Er_coords} for \Tb~ and \Er, respectively. The resulting CIF files are provided in the Supplemental Material \cite{Supplemental}.

As shown in Fig.~\ref{Fig.1}a, the crystal structure consists of alternating Ir-Al-Ge slabs and rare-earth bilayers stacked along the crystallographic $c$ axis. Within the slabs, each Ir atom is coordinated by eight Al atoms, forming a slightly distorted IrAl$_8$ cube, while the Ge atoms occupy positions above and below the Al network. The rare-earth ions are located between neighboring slabs, forming bilayers that separate successive IrAl$_4$Ge$_2$ slabs along the $c$ axis.
Since magnetism in this family is known to originate predominantly from the localized $4f$ moments of the rare-earth ions, while the transition-metal site is generally considered nonmagnetic \cite{sieve2002quaternary,huang2019physical}, the spatial arrangement of the rare-earth sublattice is expected to play a central role in determining the magnetic properties. In projection onto the $ab$ plane, the $Ln$ atoms form a perfect square network. However, the shortest $Ln$–$Ln$ distances occur between adjacent layers (approximately $3.92$ Å for Tb and $3.88$ Å for Er), indicating that the rare-earth substructure is inherently three-dimensional rather than a true two-dimensional square lattice. As a result, the magnetic coupling is not expected to be confined to the $ab$ plane, which may be relevant for understanding the complex magnetic behavior discussed below.
From a broader crystal-chemical perspective, the structure can also be viewed as an intergrowth of slabs related to simple structure types such as CeAl$_2$Ga$_2$, W, and CaF$_2$ \cite{demchenko2008quaternary}. Examples of structures built from such intergrown slabs are discussed in Ref.~\cite{parthe1993table}.

Room-temperature X-ray diffraction was performed to determine the crystallographic orientation of the single crystals. The measurements were carried out on flat facets of the plate-like crystals. As shown in Fig.\ref{Fig.1}c, the diffraction pattern exhibits only $00l$ reflections, indicating that the large flat surfaces are parallel to the $ab$ plane, with the $c$ axis oriented perpendicular to the surface. For clarity, only data for the Tb-based compound are shown, as similar results were obtained for the Er analog.

\subsection{Magnetic characterization}

\begin{figure*}[ht!]
\centering
\includegraphics[width=1\textwidth]{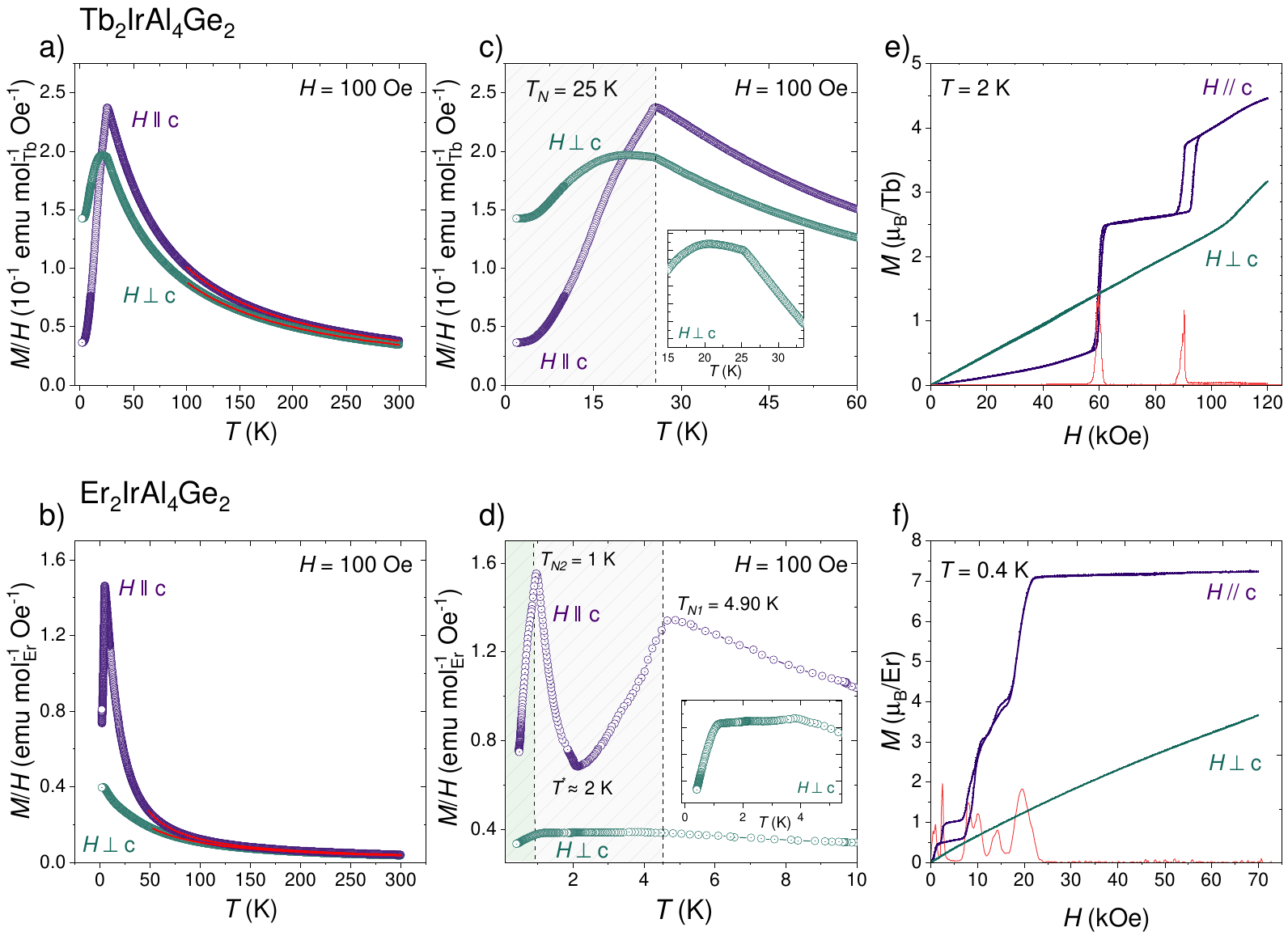}
\caption{(a,b) Field-cooled magnetization, $M/H$, for single-crystalline \Tb~ and \Er~ measured with magnetic field ($H = 100$ Oe) applied parallel and perpendicular to the crystallographic $c$ axis. Data are shown over the temperature range $1.9 \leq T \leq 300$ K, together with the Curie-Weiss fits (red lines). (c,d) Enlarged low-temperature region highlighting the magnetic transitions. For \Tb, a single antiferromagnetic transition is observed, while \Er~ exhibits three successive transitions at low temperatures. Insets in (c) and (d) highlights the magnetic behavior for $H \perp c$. (e) Isothermal magnetization of the Tb-based compound at $T = 2$ K, showing two successive metamagnetic transitions for $H \parallel c$. (f) $M(H)$ measured for \Er~ at $T = 0.4$ K for both field orientations. Multiple successive metamagnetic transitions are observed for $H \parallel c$, as emphasized by the derivative $dM/dH$ (red curve).} 
\label{fig:2}
\end{figure*}

To gain insight into the magnetic response, we first examine the low-field magnetization behavior. Figs. \ref{fig:2}a and \ref{fig:2}b display the temperature dependence of the field-cooled magnetization, $M/H$, for \Tb~ and \Er, respectively, measured at $H = 100$ Oe over the range $1.9~\mathrm{K} \le T \le 300~\mathrm{K}$. In both compounds, the magnetic response is anisotropic, with the effect being considerably more pronounced in the Er-based sample. 

At high temperatures, the magnetization follows Curie–Weiss behavior for both field orientations. The data above 50 K were therefore analyzed using a modified Curie–Weiss expression, $\chi$(\textit{T}) = $\chi_0$ + \textit{C}/(\textit{T} - $\Theta_p$), where \textit{C} is the Curie constant. For \Tb, the fits yield effective magnetic moments of $\mu_{\mathrm{eff}} = 9.77(1)\,\mu_B$ for $H \parallel c$ and $\mu_{\mathrm{eff}} = 9.61(1)\,\mu_B$ for $H \perp c$, together with Weiss temperatures of $\Theta_p = -18.3$ K and $-30.4$ K, respectively. 
Correspondingly, the \Er~ compound gives $\mu_{\mathrm{eff}} = 9.54(1)\,\mu_B$ for $H \parallel c$ and $\mu_{\mathrm{eff}} = 9.92(1)\,\mu_B$ for $H \perp c$, with Weiss temperatures of $\Theta_p = 6.5$ K and -13.8 K. The extracted effective moments are close to the theoretical free-ion values expected for Tb$^{3+}$ ($9.72\,\mu_B$) and Er$^{3+}$ ($9.58\,\mu_B$). The strong orientation dependence of the Weiss temperatures, particularly the sign change observed for \Er, reflects the pronounced single-ion magnetic anisotropy arising from crystal-electric-field (CEF) effects.

We now turn to the low-temperature behavior in the vicinity of the magnetic transitions, as shown in Figs.\ref{fig:2}c (\Tb) and \ref{fig:2}d (\Er). Only FC data are presented, as no significant differences were observed between FC and ZFC measurements. For \Tb, the magnetization reveals a single transition associated with the onset of antiferromagnetic order. For $H \parallel c$, this transition appears as a well-defined cusp-like maximum, from which the Néel temperature is determined more precisely from $\textit{d}(\textit{MT})/\textit{dT}$, giving $T_N = 25$ K. Upon rotating the field to $H \perp c$, the anomaly becomes noticeably broader and less pronounced, as highlighted in the inset. The stronger decrease of $M(T)$ for $H \parallel c$ below the magnetic transition, together with the weaker response for $H \perp c$, is consistent with the $c$ axis representing the preferred direction of magnetic moments alignment in the ordered state.

Compared to \Tb, the low-temperature magnetization of \Er~ reveals a significantly more complex magnetic response. Extending measurements to temperatures below 1.8 K reveals two successive transitions: a primary antiferromagnetic transition at $T_{N1} = 4.90$ K followed by a second anomaly at $T_{N2} = 1.0$ K, visible as an additional maximum in the $M(T)$ data. Notably, a weaker feature is also visible near $T^{*}\approx 2$ K, where the magnetization passes through a minimum followed by a subtle shoulder. While this anomaly is considerably less pronounced than those at $T_{N1}$ and $T_{N2}$ and occurs near the overlap of the conventional and $^3$He measurement ranges, it is reproducible and is further corroborated by a corresponding anomaly in the heat-capacity data discussed below. Similar to the Tb-based analog, the response for $H \perp c$ remains significantly weaker, supporting the identification of the $c$ axis as the easy magnetization direction. 

Additional insight into the magnetic anisotropy and field-induced magnetic behavior is provided by the isothermal magnetization measurements. The magnetic response of \Tb~ is presented in Fig.~\ref{fig:2}e, which shows $M(H)$ measured at $T=2$ K for two field orientations. For $H \parallel c$, two distinct metamagnetic transitions are observed near 60 and 90 kOe, accompanied by pronounced hysteresis, particularly visible for the higher-field anomaly, suggesting a first-order character. The transitions are further highlighted by the derivative $dM/dH$, calculated from the decreasing-field branch and shown as the red curve. Steep, hysteretic metamagnetic transitions of this type are consistent with spin-flip behavior commonly observed in anisotropic antiferromagnets. As established for conventional metamagnets, the field-induced transition is first order at low temperatures and becomes second order above a tricritical point \cite{bruinsma1984metamagnets}. This generic behavior provides a useful reference point for discussing the temperature evolution of the metamagnetic transitions observed here, which is considered in more detail below. Above the second transition, the magnetization continues to increase without reaching saturation up to 120 kOe, suggesting that additional field-induced magnetic phases may emerge at higher magnetic fields. In contrast, the magnetization for $H \perp c$ remains nearly linear over the entire measured field range, consistent with the weaker magnetic response observed in $M(T)$ data. We noted that the magnetization measured for fields applied within the $ab$ plane remains substantial even at the highest applied fields, indicating a relatively less magnetic anisotropy compared to that observed in \Er. A subtle change in slope is visible below approximately 110 kOe, although this feature disappears already at 3 K (not shown), suggesting that this state is only weakly stabilized. 

For \Er~ , the $M(H)$ curve measured at $T = 0.4$ K [see Fig.\ref{fig:2}f] reveals a strongly anisotropic response between the two field orientations. For $H \perp c$, the magnetization increases nearly linearly over the entire measured field range, consistent with the hard-plane character of this direction. In contrast, the $M(H)$ curve for $H \parallel c$ is significantly more complex, particularly in the low-field region. A sequence of successive field-induced transitions is observed below approximately 20 kOe, as further emphasized by pronounced anomalies in the field derivative $dM/dH$. Above the highest-field anomaly, the magnetization approaches a nearly field-independent regime, indicating proximity to a field-polarized state. Nevertheless, the magnetization remains well below the theoretical free-ion moment expected for Er$^{3+}$, pointing to a substantial influence of CEF effects that suppress the saturation moment. 

A comparison of the two compounds reveals a notable difference in the strength of the magnetic anisotropy. While both materials clearly exhibit an easy magnetization axis along the crystallographic $c$ direction, the anisotropy is considerably stronger in \Er, where the magnetization for $H \perp c$ remains strongly suppressed over the entire measured field range. In contrast, \Tb~ displays a more moderate anisotropy despite exhibiting pronounced metamagnetic behavior for $H \parallel c$. Interestingly, this behavior differs from that reported for related members of the $Ln_2M$Al$_4$Ge$_2$ family, such as \ce{Tb2NiAl4Ge2} \cite{sieve2002quaternary} and \ce{Gd2CoAl4Ge2} \cite{huang2019physical}, where easy-plane anisotropy has been observed. The change from easy-plane to easy-axis behavior upon replacing the transition-metal species suggests that the magnetic anisotropy could be sensitive to the electronic environment. Such sensitivity likely originates from modifications of the crystal-electric-field scheme together with changes in the electronic structure associated with replacing the $3d$ transition metal by the heavier $5d$ element. The stronger spin–orbit coupling and the more spatially extended $5d$ orbitals of Ir contribute to the conduction-electron states that mediate the RKKY interaction between the localized rare-earth moments, thereby altering the balance between CEF anisotropy and RKKY exchange. \cite{rau2016spin, schaffer2016recent}.

\begin{figure*}[ht!]
\centering
\includegraphics[width=1\textwidth]{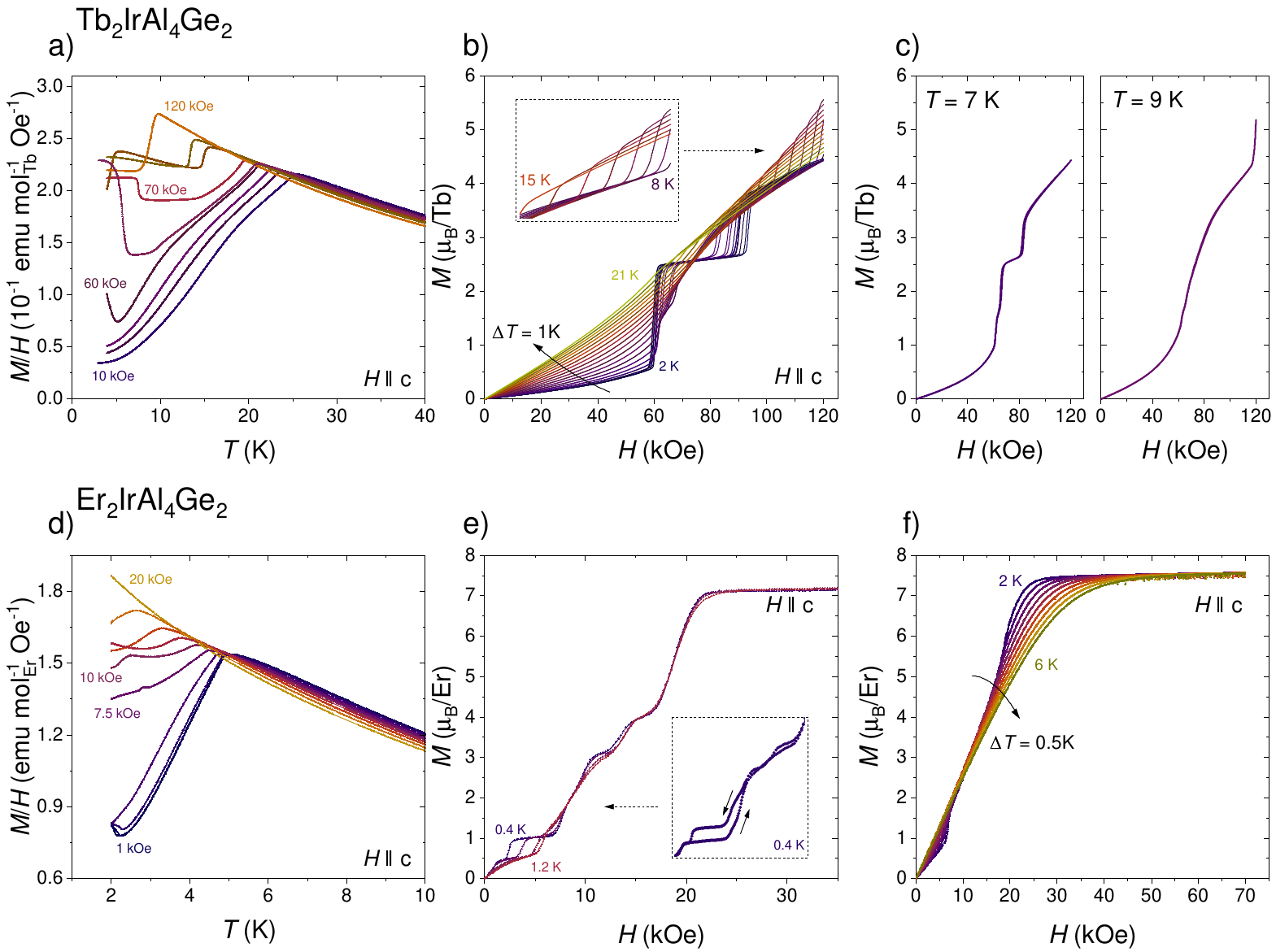}
\caption{Temperature- and field-dependent magnetization of \Tb~ and \Er~ measured with magnetic field applied parallel to the crystallographic $c$ axis. (a) Magnetization for Tb-based compound, measured under selected magnetic fields up to 120 kOe. (b) Temperature evolution of the magnetization for $H \parallel c$ between 2 and 21 K. The inset highlights the high-field anomaly emerging near 8 K. (c) Selected $M(H)$ curves measured at 7 and 9 K, illustrating additional field-induced phases.
(d) $M(T)$ data for \Er~ under magnetic fields ranging from 1 to 20 kOe. (e) Temperature evolution of the metamagnetic transitions measured between 0.4 and 1.2 K. The inset shows an enlarged view of the hysteretic low-field region at 0.4 K. (f) Magnetization measured between 2 and 6 K, showing the gradual suppression of the metamagnetic anomalies with increasing temperature.} 
\label{fig:3}
\end{figure*}

The presence of multiple field-induced transitions indicates a rich evolution of the magnetic state with both temperature and magnetic field. To examine this behavior in greater detail, temperature-dependent magnetization measurements were performed at selected magnetic fields and complemented by isothermal $M(H)$ measurements collected at various temperatures. Because the most prominent magnetic features occur for fields applied along the easy magnetization direction, only data for $H\parallel c$ are shown in Fig.\ref{fig:3}. Measurements for $H \perp c$ were also carried out (not shown here). In this orientation, however, the magnetic response shows only a weak evolution with temperature and magnetic field, and no additional anomalies were detected.

We begin with the Tb-based compound, whose comparatively simple magnetic response provides a useful starting point for understanding the field evolution of the magnetic phases. As shown in Fig.\ref{fig:3}a, the main antiferromagnetic transition gradually shifts toward lower temperatures with increasing magnetic field, consistent with the progressive suppression of the antiferromagnetic order. At the same time, the main anomaly remains clearly discernible throughout the entire investigated field range and becomes progressively sharper with increasing magnetic field. In addition to the primary transition, a field-induced anomaly emerges near 60 kOe as seen by an upturn in $M(T)$ at low temperatures. This feature is consistent with the lower-field metamagnetic transition identified in the $M(H)$ measurements at 2 K and can be followed over a finite temperature range. Initially, the anomaly shifts to higher temperatures with increasing field, but above approximately 70 kOe its evolution reverses, and it gradually moves toward lower temperatures before disappearing above 90 kOe. Although the microscopic origin of this additional feature remains unclear, its nonmonotonic field dependence suggests the presence of competing magnetic interactions that stabilize a distinct magnetic state over a limited field range.  

The temperature evolution of the two metamagnetic transitions identified in the 2 K $M(H)$ measurements is summarized in Fig.~\ref{fig:3}b. Upon warming, the two transitions exhibit markedly different temperature dependences. While the lower-field transition shifts slightly toward higher magnetic fields, the higher-field transition progressively moves toward lower fields. As a result, the field interval separating the two transitions narrows upon warming, leading to an increasingly complex magnetic response at intermediate temperatures. In particular, the $M(H)$ curve measured at 7 K, shown separately in Fig.\ref{fig:3}c, reveals the emergence of an additional field-induced phase between the two low-temperature metamagnetic transitions. This behavior suggests the existence of an intermediate pocket in the magnetic phase diagram, stabilized over a limited temperature and field range. Upon further warming, another anomaly appears at the highest measured magnetic fields near 8 K. As highlighted in the inset of Fig.\ref{fig:3}b, this transition gradually shifts toward lower magnetic fields with increasing temperature and becomes progressively broader. For clarity, the corresponding $M(H)$ curve measured at 9 K is shown separately in Fig.\ref{fig:3}c, where a pronounced anomaly is observed near 120 kOe while the magnetization remains far from saturation. These observations suggest that the third high-field metamagnetic transition is already present at the lowest temperatures, although it likely occurs beyond the experimentally accessible magnetic field range.

The magnetic response of \Er~ differs markedly from that of the Tb-based compound, reflecting the presence of multiple competing magnetic states. As shown in Fig.~\ref{fig:3}d, the primary antiferromagnetic transition gradually shifts toward lower temperatures with increasing magnetic field, while the associated anomaly becomes broader and progressively less distinct. Additional low-temperature $M(T)$ measurements performed below 1.8 K (not shown) indicate that the lowest-temperature transition at $T_{N2}$ is rapidly suppressed upon application of a magnetic field. In contrast, the intermediate feature associated with $T^*$ initially shifts toward higher temperatures and evolves into a broad maximum at intermediate fields before gradually disappearing upon further increasing the magnetic field. Above approximately 20 kOe, no clear anomalies remain in the $M(T)$ data, and the magnetization increases monotonically upon cooling, indicating that the field progressively destabilizes the low-temperature ordered phases and drives the system toward a field-polarized state.

\begin{figure}[t]
\centering
\includegraphics[width=1\columnwidth]{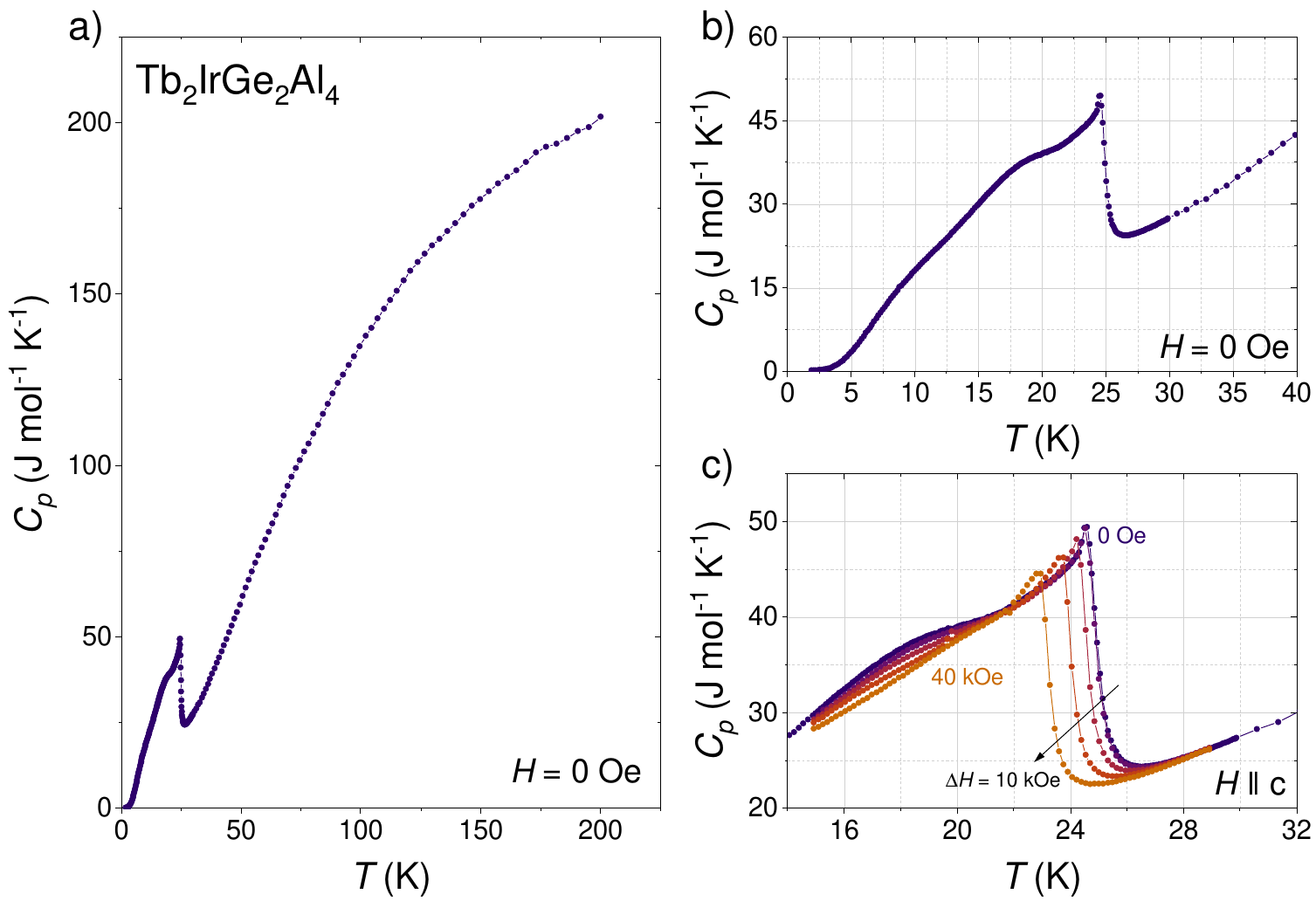}
\caption{Temperature dependence of the heat capacity $C_p(T)$ of \Tb. (a) Zero-field heat capacity measured between 1.9 and 200 K. (b) Enlarged view of the low-temperature region, showing a pronounced anomaly associated with the onset of long-range antiferromagnetic order. (c) Evolution of the low-temperature anomaly for magnetic fields up to 40 kOe applied along the crystallographic $c$ axis.}
\label{Fig.4}
\end{figure}

The complex field evolution inferred from the $M(T)$ measurements is reflected in the isothermal magnetization data shown in Figs.\ref{fig:3}e and \ref{fig:3}f. As already presented above, at 0.4 K, the $M(H)$ curve exhibits a sequence of successive metamagnetic transitions below approximately 20 kOe, as highlighted in the inset of Fig.\ref{fig:3}e. Several of these anomalies display pronounced hysteresis, indicating a first-order character of at least some of the associated phase boundaries. Upon warming, the individual transitions broaden and become less distinct. At the same time, the lowest-field anomalies shift slightly toward higher magnetic fields, revealing a nontrivial temperature evolution of the low-field magnetic phases. With further increasing temperature, the metamagnetic features gradually merge and disappear above approximately 5 K [see Fig.\ref{fig:3}f], consistent with the suppression of the ordered magnetic state above $T_{N1}$.

Taken together, the magnetization measurements reveal rich field-induced magnetic behavior in both \Tb~ and \Er. While the Tb-based compound undergoes antiferromagnetic ordering through a single transition and exhibits a sequence of successive metamagnetic transitions for $H\parallel c$, the Er-based analog displays a substantially more complex magnetic response, characterized by multiple zero-field transitions and several low-field metamagnetic anomalies. In both materials, the magnetic anisotropy and the presence of multiple field-induced transitions point to a delicate balance between exchange interactions and crystal-electric-field effects, resulting in the stabilization of several competing magnetic states.

Such behavior is similar to that reported for a number of anisotropic rare-earth magnets, including the kagome-ice compound HoAgGe with Ising spins confined to the kagome plane \cite{li2022low,zhao2020realization}, the triangular-net Ising antiferromagnet \ce{ErGa2} \cite{kurumaji2024metamagnetic,doukoure1982metamagnetism}, the square-net semimetal \ce{DyAgSb2} \cite{kusakabe2025magnetotransport}, and the quasi-one-dimensional Ising magnet \ce{TbTi3Bi4} \cite{guo20241,ortiz2024intricate}, all of which exhibit complex sequences of metamagnetic transitions and intermediate magnetic phases. Notably, although metamagnetic behavior has previously been reported in related \ce{\textit{Ln}2\textit{T}Al4Ge2} compounds such as \ce{Gd2FeAl4Ge2} \cite{he2015crystal} and \ce{Gd2CoAl4Ge2} \cite{huang2019physical}, it generally involves only a single field-induced transition. The significantly richer magnetic phase evolution observed here, therefore, highlights the distinctive character of the Ir-based members of this structural family.

\subsection{Specific heat}

To further examine the magnetic ordering, we next turn to the specific-heat measurements. Fig.\ref{Fig.4}a displays the temperature dependence of the heat capacity of \Tb~ measured between 1.9 and 200 K. The zero-field specific heat exhibits a single pronounced anomaly at low temperatures associated with the onset of long-range magnetic order, while no additional transitions are observed over the measured temperature range. A closer view of the low-temperature region is shown in Fig.\ref{Fig.4}b, where the transition temperature is found to be in very good agreement with that determined from the magnetization measurements. Below the ordering transition, the heat capacity exhibits a broad hump-like feature extending over several kelvin. Similar broad contributions are commonly observed in large-spin antiferromagnets and are consistent with the evolution of the magnetic heat capacity predicted by mean-field theory, where the anomaly becomes increasingly pronounced with increasing spin value \cite{johnston2017influence}. A quantitative comparison with these theoretical predictions would require separation of the magnetic contribution to the heat capacity. However, since no suitable nonmagnetic reference compound is available, the phonon contribution could not be reliably separated and, therefore, neither the magnetic heat capacity nor the associated magnetic entropy was determined.

The evolution of the low-temperature anomaly under magnetic fields up to 40 kOe applied along the $c$ axis is shown in Fig.\ref{Fig.4}c. As expected for an antiferromagnetic transition, the anomaly shifts toward lower temperatures with increasing field while becoming progressively broadened and suppressed. At the same time, the broad hump observed below the transition gradually weakens and eventually disappears at higher fields.

\begin{figure}[t]
\centering
\includegraphics[width=1\columnwidth]{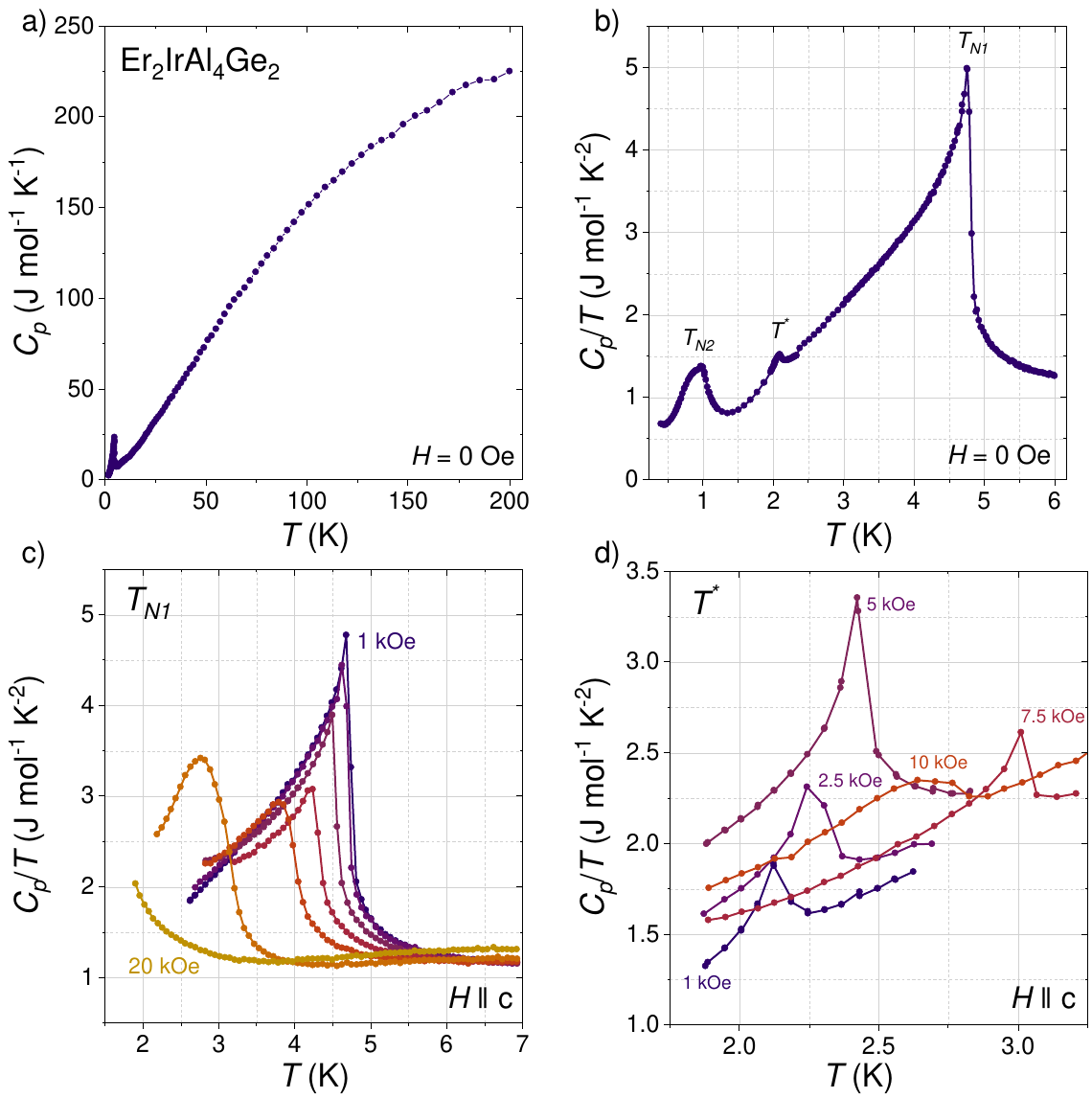}
\caption{Heat-capacity data for \Er. (a) Temperature dependence of the zero-field heat capacity $C_p(T)$. (b) Low-temperature $C_p/T$ highlighting anomalies at $T_{N1}$, $T^{*}$, and $T_{N2}$. (c) Field evolution of the anomaly at $T_{N1}$ for $H \parallel c$. (d) Enlarged view of the $T^{*}$ region under applied magnetic field.}
\label{Fig.5}
\end{figure}

The specific heat data of \Er~ is summarized in Fig.\ref{Fig.5}. Similar to the Tb-based compound, the zero-field heat capacity, measured between 1.9 and 200 K and shown in Fig.\ref{Fig.5}a, gradually approaches the classical Dulong–Petit limit of $3n$R at high temperatures, consistent with the expected phonon contribution for a system with $n=9$ atoms per formula unit. Upon cooling, the low-temperature region becomes dominated by a series of anomalies associated with magnetic ordering, reflecting the complex magnetic behavior already inferred from the magnetization measurements. A closer view of this temperature range is presented in Fig.~\ref{Fig.5}b, where three distinct features can be identified at $T_{N1}$, $T^*$, and $T_{N2}$. The sharp peak at $T_{N1} = 4.90$ K corresponds to the onset of long-range antiferromagnetic order and is in good agreement with the transition temperature determined from the magnetization measurements. The lower-temperature transitions are noticeably broader and weaker, particularly the one at $T^*$, which appeared in the magnetization data only as a subtle change in slope. This feature becomes much more evident in the heat-capacity data, confirming the presence of an additional phase transition at $T^*$. Despite the different shapes and magnitudes of these anomalies, no signatures of hysteresis or latent heat are observed, suggesting that all three transitions are second order.

The field evolution of the primary anomaly at $T_{N1}$ is shown in Fig.~\ref{Fig.5}c for selected magnetic fields up to 20 kOe applied along the $c$ axis. With increasing magnetic field, the anomaly shifts progressively toward lower temperatures and broadens slightly, consistent with the gradual suppression of the antiferromagnetic state. Interestingly, at intermediate fields near 15 kOe, the anomaly develops a sharper cusp-like character before being suppressed at higher fields. The origin of this nonmonotonic evolution remains unclear, but it may reflect changes in the magnetic correlations associated with the ordered state.

The field dependence of the intermediate anomaly at $T^*$ is presented in Fig.\ref{Fig.5}d. For low magnetic fields, the feature shifts toward higher temperatures and becomes increasingly pronounced, reaching its maximum enhancement at a field of 5 kOe. Upon further increasing the magnetic field, the anomaly continues to shift to higher temperatures, appearing near 3 K for $H = 7.5$ kOe, although its magnitude is already noticeably reduced. At still higher fields, the feature broadens substantially and gradually shifts back toward lower temperatures, as observed for 10 kOe. This non-monotonic evolution closely follows the behavior observed in the magnetization measurements, where the intermediate feature evolved into a broad maximum at intermediate fields before being suppressed at higher fields. This behavior further supports the presence of a delicate competition between nearby magnetic states, whose relative stability is strongly influenced by the applied magnetic field. 

\begin{figure*}[ht!]
\centering
\includegraphics[width=1\textwidth]{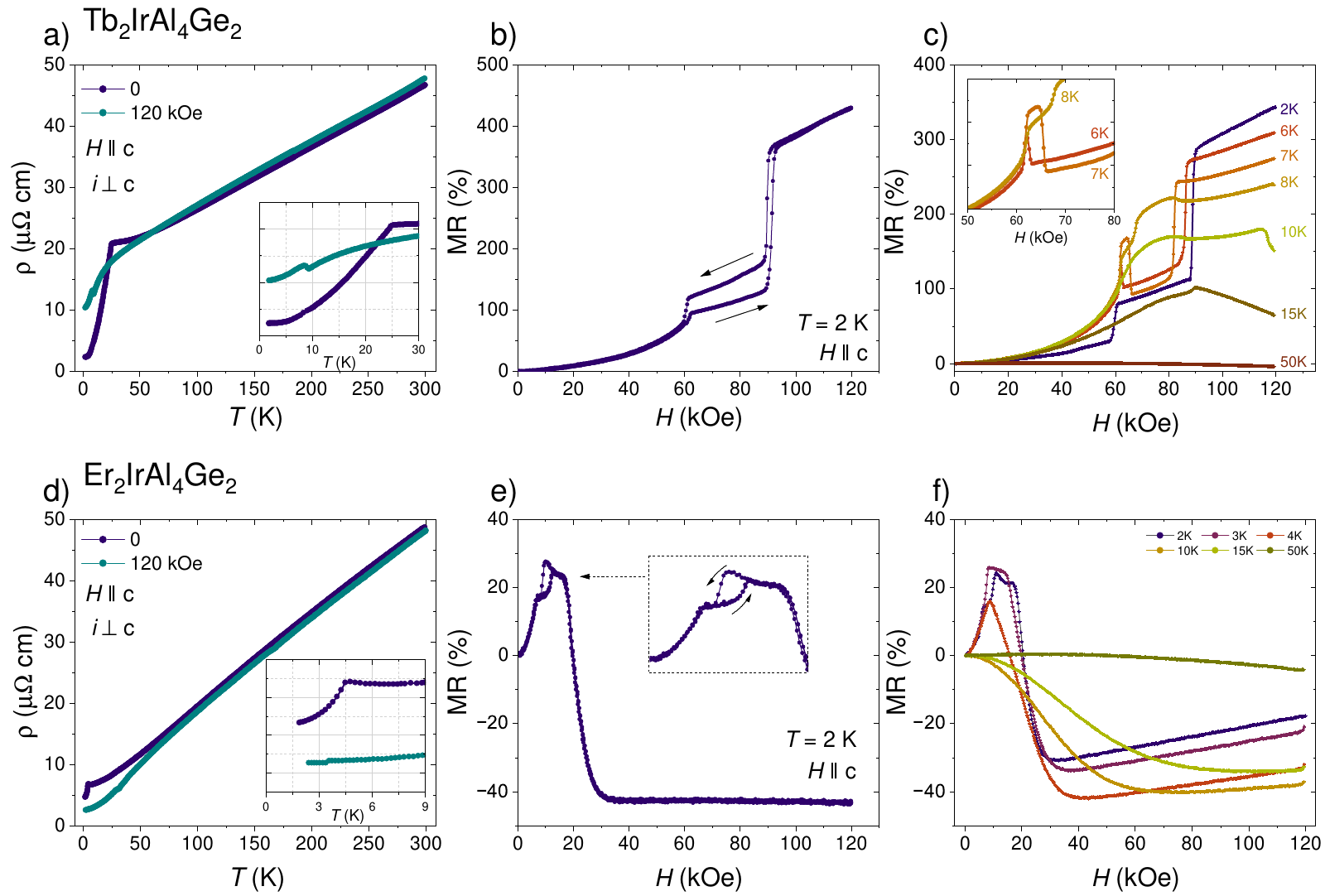}
\caption{Temperature and field dependence of the electrical resistivity and magnetoresistance of \Tb~ and \Er~ for current applied within the $ab$ plane and magnetic field applied parallel to the $c$ axis. (a,d) Electrical resistivity measured in zero field and 120 kOe. Insets show enlarged views of the low-temperature region. (b,\textcolor{BrickRed}{e}) Magnetoresistance measured at 2 K, showing the field-induced transitions and hysteretic behavior. 
(c,f) Temperature evolution of the MR, highlighting the field-induced phase transitions and their evolution with temperature.} 
\label{fig:6}
\end{figure*}

\subsection{Electronic transport}

The transport properties provide additional insight into the coupling between the conduction electrons and the magnetic order. The temperature dependence of the electrical resistivity, measured with the current applied within the $ab$ plane in zero field and in an applied field of 120 kOe, is shown in Figs.\ref{fig:6}a and \ref{fig:6}d for \Tb~ and \Er, respectively. In both compounds, the resistivity decreases monotonically with decreasing temperature, consistent with metallic behavior and the progressive reduction of electron-phonon scattering upon cooling. At low temperatures, distinct anomalies emerge in \textit{$\rho$(T)} [see insets], signaling the onset of long-range antiferromagnetic order. The transition temperatures are in good agreement with those determined from the magnetization and heat-capacity measurements. Below the transition, the resistivity decreases rapidly, indicating a substantial reduction of spin-disorder scattering following the establishment of the magnetically ordered state. This behavior may also reflect modifications of the electronic structure accompanying magnetic ordering.

The field-dependent resistivity further reflects the contrasting stability of the ordered states in the two compounds. For \Tb, the anomaly associated with antiferromagnetic ordering remains clearly visible even at 120 kOe, although it shifts to lower temperatures, reaching approximately 10 K. In contrast, the anomaly observed in \Er~ is completely suppressed at 120 kOe, resulting in a smooth metallic temperature dependence without any signatures of magnetic ordering. This is consistent with the magnetization data, which show that relatively modest magnetic fields are sufficient to destabilize the ordered state and drive the system toward a field-polarized regime. Step-like features in the 120 kOe curve are experimental artifacts associated with the measurement protocol and do not correspond to intrinsic phase transitions.

The magnetoresistance (MR), defined as [$\rho$(\textit{H})-$\rho$(0)]/$\rho$(0), of \Tb~ for $H \parallel c$ is shown in Fig.\ref{fig:6}b. To highlight the hysteretic behavior, the partial magnetoresistance loop measured at 2 K is presented as collected, without symmetrization. The magnetoresistance remains positive over the entire field range and exhibits two pronounced positive jumps associated with field-induced changes of the magnetic state, closely coinciding with the metamagnetic transitions identified in the magnetization measurements. These abrupt increases in resistivity likely reflect enhanced carrier scattering arising from distinct spin configurations stabilized across the transitions. Between the two anomalies, a plateau-like region develops and exhibits pronounced hysteresis that extends over a broad field interval. 

The temperature evolution of the magnetoresistance in \Tb~ is presented in Fig.\ref{fig:6}c. These data were symmetrized with respect to the magnetic field in order to remove the antisymmetric Hall contribution resulting from slight contact misalignment. At low temperatures, the magnetoresistance reaches approximately 350\% at 120 kOe, highlighting the strong coupling between charge transport and the underlying magnetic state. Upon warming, the low-field anomaly shifts slightly toward higher magnetic fields, while the higher-field feature gradually moves toward lower fields and becomes progressively weaker. Around 6 K, an additional feature emerges between the two low-temperature transitions, consistent with the appearance of the intermediate field-induced phase identified in the $M(H)$ measurement. This anomaly becomes particularly evident at 7 K, further supporting the existence of a narrow pocket in the $H-T$ phase diagram. At higher temperatures, the low-field anomalies are progressively suppressed, while a new feature develops at the highest magnetic fields near 8 K. As in the magnetization measurements, this anomaly shifts toward lower fields upon warming, consistent with the evolution of an additional high-field magnetic phase. Taken together, the magnetoresistance data provide an independent transport signature of the complex field-induced phase evolution in \Tb, further supporting the rich $H-T$ phase diagram established from the magnetic measurements.

The magnetoresistance for \Er~ is presented in Fig.\ref{fig:6}e. At 2 K, the low-field region exhibits several distinct anomalies together with a small but clearly discernible hysteresis (see the inset), reflecting the complex sequence of magnetic states established below 20 kOe. The magnetoresistance is initially positive and increases across the low-field anomalies. Above approximately 20 kOe, however, the MR changes sign and becomes increasingly negative as the field increases. This crossover coincides with the field range where the magnetization approaches a nearly field-independent regime, suggesting that the reduction of spin-disorder scattering becomes the dominant contribution to the transport response as the system evolves toward a field-polarized state. 

\begin{figure}[t]
\centering
\includegraphics[width=1\columnwidth]{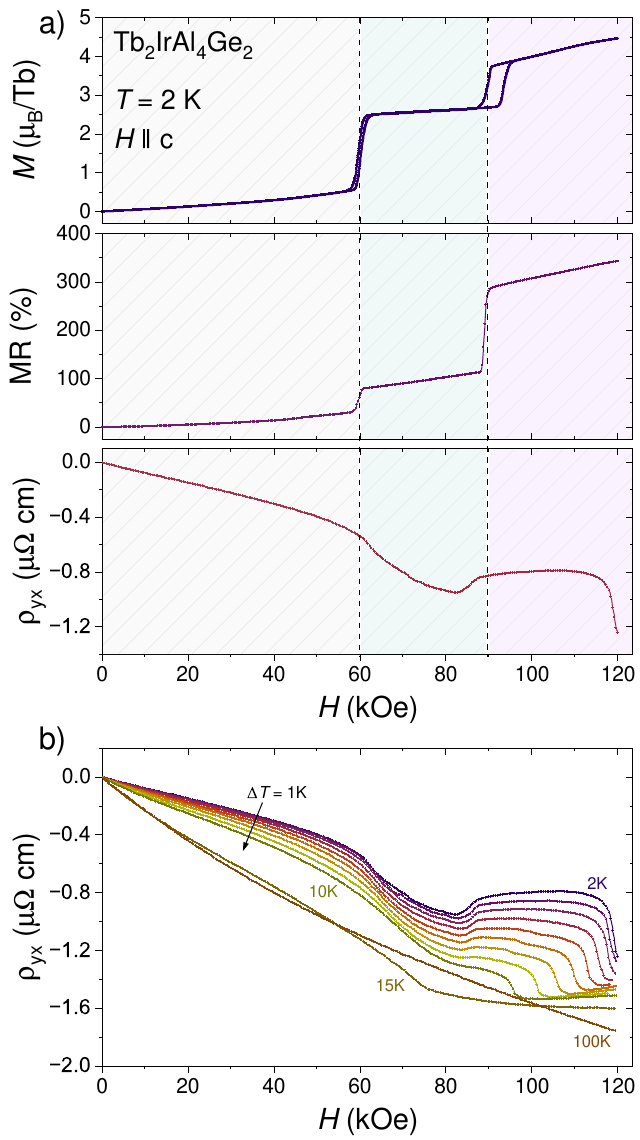}
\caption{Hall-effect measurements of \Tb~ for ($H \parallel c$). (a) Field dependence of the magnetization, magnetoresistance, and Hall resistivity measured at 2 K. (b) Temperature evolution of the Hall resistivity, showing the development and suppression of the nonlinear Hall response associated with the field-induced magnetic phases.}
\label{Fig.7}
\end{figure}

Figure \ref{fig:6}f depicts the MR data for the Er-based compound collected at different temperatures. At low temperatures, the negative magnetoresistance reaches approximately -20\% at 120 kOe. Upon warming, the low-field anomalies gradually broaden and become less distinct, while their characteristic fields shift to lower values. Above the ordering temperature, the anomalies disappear entirely, and the magnetoresistance evolves into a smooth field dependence, consistent with the suppression of the underlying magnetic order.

To further examine the influence of the field-induced magnetic phases on electronic transport, Hall-effect measurements were performed. As the Hall response is particularly sensitive to changes in the magnetic state, it provides a complementary perspective on the phase evolution inferred from the magnetization and magnetotransport data. Fig.\ref{Fig.7}a compares the field dependence of the magnetization, magnetoresistance, and Hall resistivity of \Tb~ measured at 2 K. As illustrated in the bottom panel of Fig.\ref{Fig.7}a, at low fields, the Hall resistivity $\rho_{yx}(H)$ varies nearly linearly with the magnetic field. The Hall response deviates markedly from that expected for a simple combination of ordinary and anomalous Hall contributions \cite{du2021nonlinear, nagaosa2010anomalous, zeng2006linear} upon entering the intermediate-field phase near 60 kOe, where a pronounced hump-like feature develops and persists throughout the plateau region. At the highest measured fields, an additional downward deviation emerges, suggesting that the electronic transport remains sensitive to further changes of the magnetic state in this field range. The evolution of the Hall response with temperature is shown in Fig.~\ref{Fig.7}b. The hump-like contribution associated with the intermediate-field phase is most pronounced at low temperatures and weakens upon warming. Simultaneously, the high-field anomaly shifts toward lower magnetic fields, tracking the evolution of the corresponding phase boundary established from the magnetization and magnetotransport measurements. Above approximately 15 K, the nonlinear contribution becomes strongly reduced, and by 100 K the Hall resistivity is approximately linear over the entire measured field range. For completeness, analogous Hall-effect measurements were also performed for \Er~ at 2 K. In contrast to \Tb, the Hall resistivity remains approximately linear over the entire measured field range, and no anomalies associated with the magnetic transitions were detected (data not shown).

The close correspondence between the Hall anomalies and the field-induced magnetic phases indicates that the Hall response is strongly influenced by the underlying magnetic state. In particular, the development of the hump-like contribution within the intermediate-field phase suggests a substantial modification of the electronic transport upon entering this region of the phase diagram. Similar nonlinear Hall responses have been reported in a variety of correlated magnetic systems, where they are often discussed in the context of unconventional magnetic textures and Berry-phase-related effects \cite{liu2023topological,wang2021field, wang2023anomalous, ueda2012topological, rout2019field, shiomi2012hall, liu2017transition, surgers2014large}. However, comparable features have also been observed in materials with predominantly collinear or Ising-like magnetic moments, where they have been attributed to mechanisms such as changes in carrier mobility, magnetic scattering, or modifications of the electronic structure across magnetic phase transitions \cite{kurumaji2024metamagnetic, kimbell2022challenges}. At present, the microscopic origin of the Hall anomaly in \Tb~ remains unclear. Nevertheless, its strong correlation with the field-induced phase evolution indicates that it is intimately connected to the magnetic degrees of freedom of the system.

\subsection{Magnetic phase diagrams}

\begin{figure}[t]
\centering
\includegraphics[width=1\columnwidth]{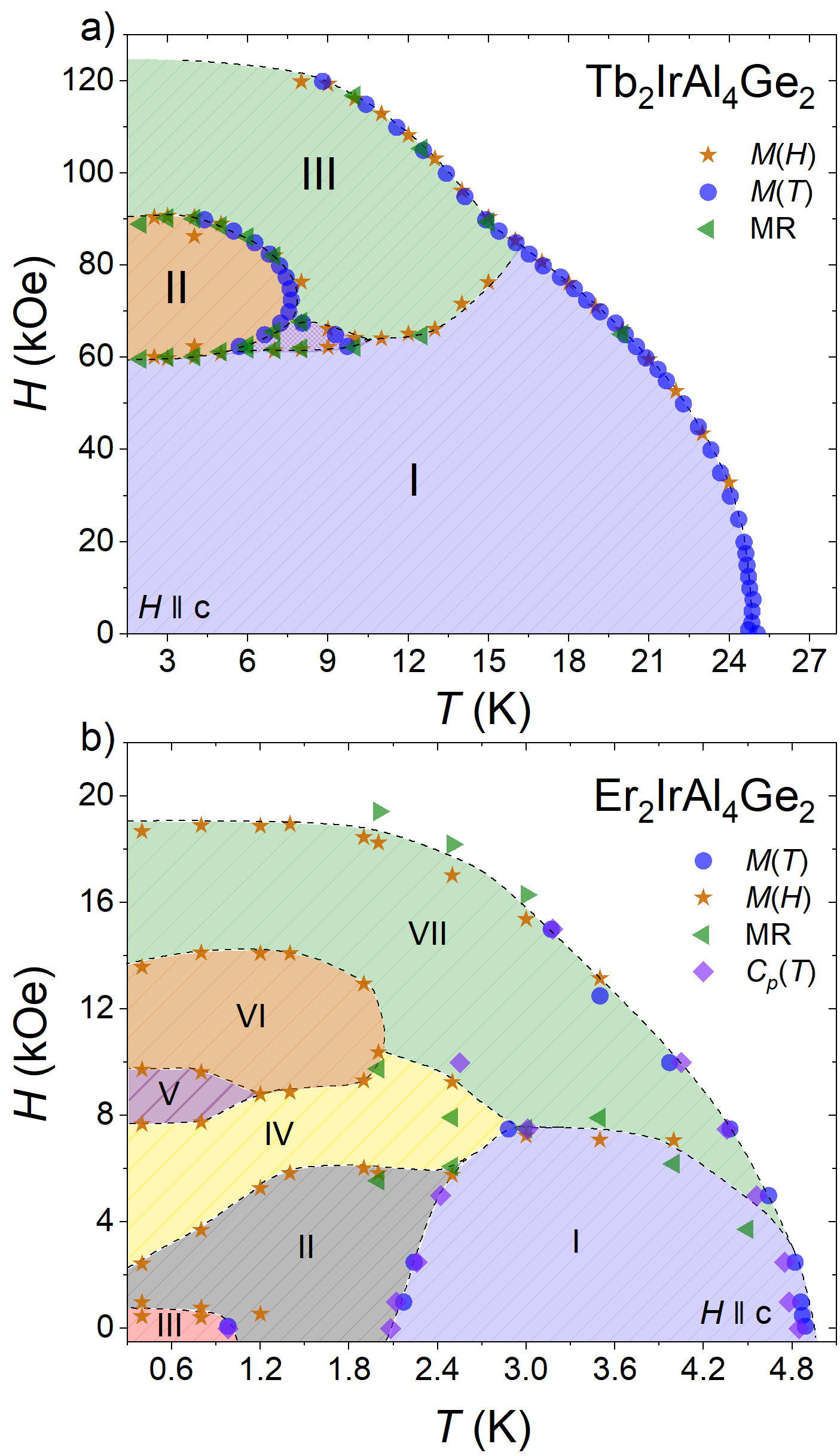}
\caption{Magnetic field–temperature phase diagrams of (a) \Tb~ and (b) \Er~ for $H \parallel c$. Symbols indicate experimentally determined phase boundaries; colored regions and lines are guides to the eye.}
\label{Fig.8}
\end{figure}

The magnetic phase diagrams for \Tb~ and \Er~ with the magnetic field applied along the easy magnetization direction ($H \parallel c$) are summarized in Fig.\ref{Fig.8}. The phase boundaries were determined from anomalies identified in the temperature-dependent magnetization, isothermal magnetization, heat-capacity, and magnetoresistance measurements. The colored regions and lines are intended primarily to guide the eye. In the absence of microscopic information about the magnetic structures, the proposed phases should be viewed as experimentally defined magnetic regions whose precise nature remains to be determined.

For \Tb~ [Fig.\ref{Fig.8}a], long-range antiferromagnetic order emerges below $T_N = 25$ K and remains stable throughout the investigated field range. At the lowest temperatures, the application of magnetic field along the $c$ axis drives the system through a sequence of successive metamagnetic transitions that separate three distinct magnetic regions, labeled I–III. The boundaries separating phases I and II, as well as phases II and III, are accompanied by pronounced hysteresis in both the magnetization and magnetoresistance measurements, indicating that these transitions possess a first-order character. The Hall-effect measurements further show that the strongest nonlinear Hall contribution develops within phase II, pointing to a substantial modification of the electronic transport upon entering this field-induced state. At intermediate temperatures, an additional pocket-like phase is stabilized within a narrow range of temperature and magnetic field near the convergence of the major phase boundaries, highlighting the close competition between several field-induced magnetic states. Finally, the evolution of the magnetization, magnetoresistance, and Hall response at the highest accessible fields suggests the onset of an additional field-induced transition above 120 kOe. As a result, the field-polarized state is likely reached only at magnetic fields beyond the experimental range investigated in the present work.

The phase diagram of \Er~ is considerably more intricate than that of its Tb-based counterpart. Already in zero magnetic field, three successive phase transitions are observed at $T_{N1}$, $T^{*}$, and $T_{N2}$, corresponding to a complex sequence of magnetic phases upon cooling. The application of a magnetic field affects these transitions in markedly different ways. While the primary ordering temperature $T_{N1}$ is progressively suppressed, the intermediate transition at $T^{*}$ initially shifts to higher temperatures before evolving into a broad maximum and eventually being fully suppressed. The phase diagram becomes even richer below approximately 1 K, where multiple field-induced transitions emerge. At the lowest temperatures, a sequence of metamagnetic transitions below 20 kOe gives rise to at least six distinct magnetic regions. In contrast to \Tb, the magnetic phases of \Er~ are stabilized over a considerably narrower region of the $H-T$ phase space, resulting in a significantly larger number of closely spaced phase boundaries.

Taken together, the phase diagrams of both compounds indicate a delicate balance between competing magnetic interactions. The stabilization of multiple field-induced phases likely reflects the interplay between long-range RKKY exchange interactions, which mediate coupling between the localized $4f$ moments, and crystal-electric-field effects that impose strong magnetic anisotropy on the system. While this competition gives rise to a rich sequence of magnetic phases in \Tb, the considerably greater complexity of the \Er~ phase diagram suggests an even more delicate balance between these energy scales, leading to the stabilization of a larger number of closely competing magnetic states. Although the present thermodynamic, magnetic, and transport measurements provide a comprehensive picture of the phase evolution, the microscopic nature of the individual phases remains unresolved. Further investigations, particularly neutron diffraction measurements under applied magnetic field, will be essential for determining the corresponding magnetic structures and for clarifying the mechanisms responsible for their stabilization.

\section{Summary}
In summary, single crystals of \Tb~ and \Er~, the first Ir-based representatives of the $Ln_2M$Al$_4$Ge$_2$ family, have been successfully synthesized and characterized. Both compounds crystallize in the tetragonal \ce{Tb2NiAl4Ge2}-type structure and exhibit pronounced magnetic anisotropy with the crystallographic $c$ axis serving as the easy magnetization direction.

Despite their structural similarity, the two compounds exhibit markedly different magnetic behaviors. \Tb~ undergoes a single antiferromagnetic transition at the relatively high temperature of $T_{N} = 25$ K, followed by a sequence of field-induced phases that remain stable to high magnetic fields. By contrast, \Er~ exhibits three successive magnetic transitions below 5 K and a substantially more intricate low-temperature phase diagram comprising several closely competing magnetic states. While both compounds possess an easy-axis anisotropy, the anisotropy is considerably stronger in \Er~ than in \Tb. This behavior contrasts with that reported for the related compounds, such as \ce{Tb2NiAl4Ge2} \cite{sieve2002quaternary}, where easy-plane anisotropy has been observed, suggesting that the magnetic anisotropy within the $Ln_2M$Al$_4$Ge$_2$ family may be tuned by the choice of transition-metal element. In both materials, the magnetic phase boundaries are clearly reflected in the magnetotransport response, demonstrating a strong coupling between charge transport and magnetic order.

The present results demonstrate that incorporating the heavier $5d$ transition metal Ir substantially enriches the magnetic behavior within the $Ln_2M$Al$_4$Ge$_2$ structure type, giving rise to complex magnetic phase diagrams and multiple field-induced states. More broadly, they establish Ir-based aluminogermanides as a promising platform for investigating the interplay between magnetic anisotropy, competing exchange interactions, and electronic transport. The successful stabilization of Ir within this structure type further suggests that a wider range of $4d$- and $5d$-based compounds may be accessible, thereby considerably expanding the chemical landscape of this family. Future studies combining microscopic probes, such as neutron diffraction, with spectroscopic techniques sensitive to crystal-electric-field excitations and magnetic dynamics will help elucidate the nature of the competing magnetic phases and the mechanisms governing their evolution under applied magnetic field.

\section{Acknowledgment}
This work was supported by the U.S. Department of Energy, Office of Science, Basic Energy Sciences, Materials Sciences and Engineering Division. 

\section{Data Availability}
The data that support the findings of this article are openly available \cite{Data}.


\begin{thebibliography}{59}%
\makeatletter
\providecommand \@ifxundefined [1]{%
 \@ifx{#1\undefined}
}%
\providecommand \@ifnum [1]{%
 \ifnum #1\expandafter \@firstoftwo
 \else \expandafter \@secondoftwo
 \fi
}%
\providecommand \@ifx [1]{%
 \ifx #1\expandafter \@firstoftwo
 \else \expandafter \@secondoftwo
 \fi
}%
\providecommand \natexlab [1]{#1}%
\providecommand \enquote  [1]{``#1''}%
\providecommand \bibnamefont  [1]{#1}%
\providecommand \bibfnamefont [1]{#1}%
\providecommand \citenamefont [1]{#1}%
\providecommand \href@noop [0]{\@secondoftwo}%
\providecommand \href [0]{\begingroup \@sanitize@url \@href}%
\providecommand \@href[1]{\@@startlink{#1}\@@href}%
\providecommand \@@href[1]{\endgroup#1\@@endlink}%
\providecommand \@sanitize@url [0]{\catcode `\\12\catcode `\$12\catcode `\&12\catcode `\#12\catcode `\^12\catcode `\_12\catcode `\%12\relax}%
\providecommand \@@startlink[1]{}%
\providecommand \@@endlink[0]{}%
\providecommand \url  [0]{\begingroup\@sanitize@url \@url }%
\providecommand \@url [1]{\endgroup\@href {#1}{\urlprefix }}%
\providecommand \urlprefix  [0]{URL }%
\providecommand \Eprint [0]{\href }%
\providecommand \doibase [0]{https://doi.org/}%
\providecommand \selectlanguage [0]{\@gobble}%
\providecommand \bibinfo  [0]{\@secondoftwo}%
\providecommand \bibfield  [0]{\@secondoftwo}%
\providecommand \translation [1]{[#1]}%
\providecommand \BibitemOpen [0]{}%
\providecommand \bibitemStop [0]{}%
\providecommand \bibitemNoStop [0]{.\EOS\space}%
\providecommand \EOS [0]{\spacefactor3000\relax}%
\providecommand \BibitemShut  [1]{\csname bibitem#1\endcsname}%
\let\auto@bib@innerbib\@empty
\bibitem [{\citenamefont {Ruderman}\ and\ \citenamefont {Kittel}(1954)}]{ruderman1954indirect}%
  \BibitemOpen
  \bibfield  {author} {\bibinfo {author} {\bibfnamefont {M.~A.}\ \bibnamefont {Ruderman}}\ and\ \bibinfo {author} {\bibfnamefont {C.}~\bibnamefont {Kittel}},\ }\bibfield  {title} {\bibinfo {title} {Indirect exchange coupling of nuclear magnetic moments by conduction electrons},\ }\href@noop {} {\bibfield  {journal} {\bibinfo  {journal} {Physical Review}\ }\textbf {\bibinfo {volume} {96}},\ \bibinfo {pages} {99} (\bibinfo {year} {1954})}\BibitemShut {NoStop}%
\bibitem [{\citenamefont {Van~Vleck}(1962)}]{van1962note}%
  \BibitemOpen
  \bibfield  {author} {\bibinfo {author} {\bibfnamefont {J.}~\bibnamefont {Van~Vleck}},\ }\bibfield  {title} {\bibinfo {title} {Note on the interactions between the spins of magnetic ions or nuclei in metals},\ }\href@noop {} {\bibfield  {journal} {\bibinfo  {journal} {Reviews of Modern Physics}\ }\textbf {\bibinfo {volume} {34}},\ \bibinfo {pages} {681} (\bibinfo {year} {1962})}\BibitemShut {NoStop}%
\bibitem [{\citenamefont {Kasuya}(1956)}]{kasuya1956theory}%
  \BibitemOpen
  \bibfield  {author} {\bibinfo {author} {\bibfnamefont {T.}~\bibnamefont {Kasuya}},\ }\bibfield  {title} {\bibinfo {title} {A theory of metallic ferro-and antiferromagnetism on {Zener's} model},\ }\href@noop {} {\bibfield  {journal} {\bibinfo  {journal} {Progress of Theoretical Physics}\ }\textbf {\bibinfo {volume} {16}},\ \bibinfo {pages} {45} (\bibinfo {year} {1956})}\BibitemShut {NoStop}%
\bibitem [{\citenamefont {Xie}\ \emph {et~al.}(2021)\citenamefont {Xie}, \citenamefont {Luo}, \citenamefont {Su}, \citenamefont {Zheng}, \citenamefont {Nie}, \citenamefont {Smidman}, \citenamefont {Takabatake},\ and\ \citenamefont {Yuan}}]{xie2021complex}%
  \BibitemOpen
  \bibfield  {author} {\bibinfo {author} {\bibfnamefont {W.}~\bibnamefont {Xie}}, \bibinfo {author} {\bibfnamefont {S.}~\bibnamefont {Luo}}, \bibinfo {author} {\bibfnamefont {H.}~\bibnamefont {Su}}, \bibinfo {author} {\bibfnamefont {X.}~\bibnamefont {Zheng}}, \bibinfo {author} {\bibfnamefont {Z.}~\bibnamefont {Nie}}, \bibinfo {author} {\bibfnamefont {M.}~\bibnamefont {Smidman}}, \bibinfo {author} {\bibfnamefont {T.}~\bibnamefont {Takabatake}},\ and\ \bibinfo {author} {\bibfnamefont {H.}~\bibnamefont {Yuan}},\ }\bibfield  {title} {\bibinfo {title} {Complex magnetic phase diagram in noncentrosymmetric \ce{EuPtAs}},\ }\href@noop {} {\bibfield  {journal} {\bibinfo  {journal} {Physical Review B}\ }\textbf {\bibinfo {volume} {104}},\ \bibinfo {pages} {174425} (\bibinfo {year} {2021})}\BibitemShut {NoStop}%
\bibitem [{\citenamefont {Wu}\ \emph {et~al.}(2024)\citenamefont {Wu}, \citenamefont {Zhang}, \citenamefont {Ren}, \citenamefont {Zhang}, \citenamefont {Gao}, \citenamefont {An}, \citenamefont {Li},\ and\ \citenamefont {Zhang}}]{wu2024transition}%
  \BibitemOpen
  \bibfield  {author} {\bibinfo {author} {\bibfnamefont {H.}~\bibnamefont {Wu}}, \bibinfo {author} {\bibfnamefont {L.}~\bibnamefont {Zhang}}, \bibinfo {author} {\bibfnamefont {W.}~\bibnamefont {Ren}}, \bibinfo {author} {\bibfnamefont {Q.}~\bibnamefont {Zhang}}, \bibinfo {author} {\bibfnamefont {F.}~\bibnamefont {Gao}}, \bibinfo {author} {\bibfnamefont {M.}~\bibnamefont {An}}, \bibinfo {author} {\bibfnamefont {B.}~\bibnamefont {Li}},\ and\ \bibinfo {author} {\bibfnamefont {Z.}~\bibnamefont {Zhang}},\ }\bibfield  {title} {\bibinfo {title} {Transition from an incommensurate spin density wave to a commensurate magnetic order in a triangular lattice compound \ce{Ho2PdAl6Ge4}},\ }\href@noop {} {\bibfield  {journal} {\bibinfo  {journal} {Journal of Alloys and Compounds}\ }\textbf {\bibinfo {volume} {998}},\ \bibinfo {pages} {174975} (\bibinfo {year} {2024})}\BibitemShut {NoStop}%
\bibitem [{\citenamefont {Wu}\ \emph {et~al.}(2026)\citenamefont {Wu}, \citenamefont {Ren}, \citenamefont {Wang}, \citenamefont {Cong}, \citenamefont {Ge}, \citenamefont {Wu}, \citenamefont {Gao}, \citenamefont {An}, \citenamefont {Li},\ and\ \citenamefont {Zhang}}]{wu2026possible}%
  \BibitemOpen
  \bibfield  {author} {\bibinfo {author} {\bibfnamefont {H.}~\bibnamefont {Wu}}, \bibinfo {author} {\bibfnamefont {W.}~\bibnamefont {Ren}}, \bibinfo {author} {\bibfnamefont {C.-W.}\ \bibnamefont {Wang}}, \bibinfo {author} {\bibfnamefont {M.}~\bibnamefont {Cong}}, \bibinfo {author} {\bibfnamefont {H.}~\bibnamefont {Ge}}, \bibinfo {author} {\bibfnamefont {L.}~\bibnamefont {Wu}}, \bibinfo {author} {\bibfnamefont {F.}~\bibnamefont {Gao}}, \bibinfo {author} {\bibfnamefont {M.}~\bibnamefont {An}}, \bibinfo {author} {\bibfnamefont {B.}~\bibnamefont {Li}},\ and\ \bibinfo {author} {\bibfnamefont {Z.}~\bibnamefont {Zhang}},\ }\bibfield  {title} {\bibinfo {title} {{Possible coexistence of non-Fermi liquid behavior and incommensurate spin density wave in the triangular lattice compound \ce{TbAl2Si2}}},\ }\href@noop {} {\bibfield  {journal} {\bibinfo  {journal} {Physical Review B}\ }\textbf {\bibinfo {volume} {113}},\ \bibinfo {pages} {054414} (\bibinfo {year} {2026})}\BibitemShut {NoStop}%
\bibitem [{\citenamefont {Gignoux}\ and\ \citenamefont {Schmitt}(1993)}]{gignoux1993competition}%
  \BibitemOpen
  \bibfield  {author} {\bibinfo {author} {\bibfnamefont {D.}~\bibnamefont {Gignoux}}\ and\ \bibinfo {author} {\bibfnamefont {D.}~\bibnamefont {Schmitt}},\ }\bibfield  {title} {\bibinfo {title} {{Competition between commensurate and incommensurate phases in rare-earth systems: Effects on H-T magnetic phase diagrams}},\ }\href@noop {} {\bibfield  {journal} {\bibinfo  {journal} {Physical Review B}\ }\textbf {\bibinfo {volume} {48}},\ \bibinfo {pages} {12682} (\bibinfo {year} {1993})}\BibitemShut {NoStop}%
\bibitem [{\citenamefont {Feng}\ \emph {et~al.}(2026)\citenamefont {Feng}, \citenamefont {Anderson}, \citenamefont {Rosmus}, \citenamefont {Antezak}, \citenamefont {Farago}, \citenamefont {Frantzeskakis}, \citenamefont {Fortuna}, \citenamefont {Santander-Syro}, \citenamefont {McCandless}, \citenamefont {Chan} \emph {et~al.}}]{feng2026complex}%
  \BibitemOpen
  \bibfield  {author} {\bibinfo {author} {\bibfnamefont {K.}~\bibnamefont {Feng}}, \bibinfo {author} {\bibfnamefont {M.~G.}\ \bibnamefont {Anderson}}, \bibinfo {author} {\bibfnamefont {M.}~\bibnamefont {Rosmus}}, \bibinfo {author} {\bibfnamefont {A.}~\bibnamefont {Antezak}}, \bibinfo {author} {\bibfnamefont {C.}~\bibnamefont {Farago}}, \bibinfo {author} {\bibfnamefont {E.}~\bibnamefont {Frantzeskakis}}, \bibinfo {author} {\bibfnamefont {F.}~\bibnamefont {Fortuna}}, \bibinfo {author} {\bibfnamefont {A.~F.}\ \bibnamefont {Santander-Syro}}, \bibinfo {author} {\bibfnamefont {G.~T.}\ \bibnamefont {McCandless}}, \bibinfo {author} {\bibfnamefont {J.~Y.}\ \bibnamefont {Chan}}, \emph {et~al.},\ }\bibfield  {title} {\bibinfo {title} {Complex magnetic transitions in the triangular net compound \ce{GdAl2Ge2}},\ }\href@noop {} {\bibfield  {journal} {\bibinfo  {journal} {Physical Review B}\ }\textbf {\bibinfo {volume} {113}},\ \bibinfo {pages} {144430} (\bibinfo {year} {2026})}\BibitemShut {NoStop}%
\bibitem [{\citenamefont {Khan}\ \emph {et~al.}(2019)\citenamefont {Khan}, \citenamefont {Zhang}, \citenamefont {Bao}, \citenamefont {Fishman}, \citenamefont {Botana}, \citenamefont {Choi}, \citenamefont {Fabbris}, \citenamefont {Haskel}, \citenamefont {Singleton},\ and\ \citenamefont {Mitchell}}]{khan2019steplike}%
  \BibitemOpen
  \bibfield  {author} {\bibinfo {author} {\bibfnamefont {M.~A.}\ \bibnamefont {Khan}}, \bibinfo {author} {\bibfnamefont {Q.}~\bibnamefont {Zhang}}, \bibinfo {author} {\bibfnamefont {J.-K.}\ \bibnamefont {Bao}}, \bibinfo {author} {\bibfnamefont {R.~S.}\ \bibnamefont {Fishman}}, \bibinfo {author} {\bibfnamefont {A.~S.}\ \bibnamefont {Botana}}, \bibinfo {author} {\bibfnamefont {Y.}~\bibnamefont {Choi}}, \bibinfo {author} {\bibfnamefont {G.}~\bibnamefont {Fabbris}}, \bibinfo {author} {\bibfnamefont {D.}~\bibnamefont {Haskel}}, \bibinfo {author} {\bibfnamefont {J.}~\bibnamefont {Singleton}},\ and\ \bibinfo {author} {\bibfnamefont {J.~F.}\ \bibnamefont {Mitchell}},\ }\bibfield  {title} {\bibinfo {title} {Steplike metamagnetic transitions in a honeycomb lattice antiferromagnet \ce{Tb2Ir3Ga9}},\ }\href@noop {} {\bibfield  {journal} {\bibinfo  {journal} {Physical Review Materials}\ }\textbf {\bibinfo {volume} {3}},\ \bibinfo {pages} {114411} (\bibinfo {year} {2019})}\BibitemShut {NoStop}%
\bibitem [{\citenamefont {Ram}\ \emph {et~al.}(2023)\citenamefont {Ram}, \citenamefont {Singh}, \citenamefont {Hooda}, \citenamefont {Singh}, \citenamefont {Kanchana}, \citenamefont {Kaczorowski},\ and\ \citenamefont {Hossain}}]{ram2023multiple}%
  \BibitemOpen
  \bibfield  {author} {\bibinfo {author} {\bibfnamefont {D.}~\bibnamefont {Ram}}, \bibinfo {author} {\bibfnamefont {J.}~\bibnamefont {Singh}}, \bibinfo {author} {\bibfnamefont {M.}~\bibnamefont {Hooda}}, \bibinfo {author} {\bibfnamefont {K.}~\bibnamefont {Singh}}, \bibinfo {author} {\bibfnamefont {V.}~\bibnamefont {Kanchana}}, \bibinfo {author} {\bibfnamefont {D.}~\bibnamefont {Kaczorowski}},\ and\ \bibinfo {author} {\bibfnamefont {Z.}~\bibnamefont {Hossain}},\ }\bibfield  {title} {\bibinfo {title} {Multiple magnetic transitions, metamagnetism, and large magnetoresistance in \ce{GdAuGe} single crystals},\ }\href@noop {} {\bibfield  {journal} {\bibinfo  {journal} {Physical Review B}\ }\textbf {\bibinfo {volume} {108}},\ \bibinfo {pages} {235107} (\bibinfo {year} {2023})}\BibitemShut {NoStop}%
\bibitem [{\citenamefont {Sharma}\ \emph {et~al.}(2022)\citenamefont {Sharma}, \citenamefont {Dan}, \citenamefont {Nandi}, \citenamefont {Dwari}, \citenamefont {Maity}, \citenamefont {Sanki}, \citenamefont {Kulkarni},\ and\ \citenamefont {Thamizhavel}}]{sharma2022complex}%
  \BibitemOpen
  \bibfield  {author} {\bibinfo {author} {\bibfnamefont {V.}~\bibnamefont {Sharma}}, \bibinfo {author} {\bibfnamefont {S.}~\bibnamefont {Dan}}, \bibinfo {author} {\bibfnamefont {S.}~\bibnamefont {Nandi}}, \bibinfo {author} {\bibfnamefont {G.}~\bibnamefont {Dwari}}, \bibinfo {author} {\bibfnamefont {B.~B.}\ \bibnamefont {Maity}}, \bibinfo {author} {\bibfnamefont {S.}~\bibnamefont {Sanki}}, \bibinfo {author} {\bibfnamefont {R.}~\bibnamefont {Kulkarni}},\ and\ \bibinfo {author} {\bibfnamefont {A.}~\bibnamefont {Thamizhavel}},\ }\bibfield  {title} {\bibinfo {title} {Complex magnetic behavior in \ce{Ho2Re3Si5} single crystal},\ }\href@noop {} {\bibfield  {journal} {\bibinfo  {journal} {Physical Review Materials}\ }\textbf {\bibinfo {volume} {6}},\ \bibinfo {pages} {124406} (\bibinfo {year} {2022})}\BibitemShut {NoStop}%
\bibitem [{\citenamefont {Gam{\.z}a}\ \emph {et~al.}(2019)\citenamefont {Gam{\.z}a}, \citenamefont {Schnelle}, \citenamefont {Rosner}, \citenamefont {Ackerbauer}, \citenamefont {Grin},\ and\ \citenamefont {Leithe-Jasper}}]{gamza2019complex}%
  \BibitemOpen
  \bibfield  {author} {\bibinfo {author} {\bibfnamefont {M.~B.}\ \bibnamefont {Gam{\.z}a}}, \bibinfo {author} {\bibfnamefont {W.}~\bibnamefont {Schnelle}}, \bibinfo {author} {\bibfnamefont {H.}~\bibnamefont {Rosner}}, \bibinfo {author} {\bibfnamefont {S.-V.}\ \bibnamefont {Ackerbauer}}, \bibinfo {author} {\bibfnamefont {Y.}~\bibnamefont {Grin}},\ and\ \bibinfo {author} {\bibfnamefont {A.}~\bibnamefont {Leithe-Jasper}},\ }\bibfield  {title} {\bibinfo {title} {Complex magnetic phase diagram of metamagnetic \ce{MnPtSi}},\ }\href@noop {} {\bibfield  {journal} {\bibinfo  {journal} {Physical Review B}\ }\textbf {\bibinfo {volume} {100}},\ \bibinfo {pages} {014423} (\bibinfo {year} {2019})}\BibitemShut {NoStop}%
\bibitem [{\citenamefont {Yang}\ \emph {et~al.}(2025)\citenamefont {Yang}, \citenamefont {Pan}, \citenamefont {Yang}, \citenamefont {Chu},\ and\ \citenamefont {Liu}}]{yang2025two}%
  \BibitemOpen
  \bibfield  {author} {\bibinfo {author} {\bibfnamefont {X.}~\bibnamefont {Yang}}, \bibinfo {author} {\bibfnamefont {J.}~\bibnamefont {Pan}}, \bibinfo {author} {\bibfnamefont {M.}~\bibnamefont {Yang}}, \bibinfo {author} {\bibfnamefont {D.}~\bibnamefont {Chu}},\ and\ \bibinfo {author} {\bibfnamefont {S.}~\bibnamefont {Liu}},\ }\bibfield  {title} {\bibinfo {title} {{Two-dimensional Critical behavior, complex magnetic phase diagram, and anisotropic magnetic entropy change of a two-dimensional kagome magnet \ce{GdTi3Bi4}}},\ }\href@noop {} {\bibfield  {journal} {\bibinfo  {journal} {Journal of Magnetism and Magnetic Materials}\ }\textbf {\bibinfo {volume} {621}},\ \bibinfo {pages} {172916} (\bibinfo {year} {2025})}\BibitemShut {NoStop}%
\bibitem [{\citenamefont {Feng}\ \emph {et~al.}(2022)\citenamefont {Feng}, \citenamefont {Leahy}, \citenamefont {Oladehin}, \citenamefont {Wei}, \citenamefont {Lee},\ and\ \citenamefont {Baumbach}}]{feng2022magnetic}%
  \BibitemOpen
  \bibfield  {author} {\bibinfo {author} {\bibfnamefont {K.}~\bibnamefont {Feng}}, \bibinfo {author} {\bibfnamefont {I.~A.}\ \bibnamefont {Leahy}}, \bibinfo {author} {\bibfnamefont {O.}~\bibnamefont {Oladehin}}, \bibinfo {author} {\bibfnamefont {K.}~\bibnamefont {Wei}}, \bibinfo {author} {\bibfnamefont {M.}~\bibnamefont {Lee}},\ and\ \bibinfo {author} {\bibfnamefont {R.}~\bibnamefont {Baumbach}},\ }\bibfield  {title} {\bibinfo {title} {Magnetic ordering in \ce{GdAuAl4Ge2} and \ce{TbAuAl4Ge2}: Layered compounds with triangular lanthanide nets},\ }\href@noop {} {\bibfield  {journal} {\bibinfo  {journal} {Journal of Magnetism and Magnetic Materials}\ }\textbf {\bibinfo {volume} {564}},\ \bibinfo {pages} {170006} (\bibinfo {year} {2022})}\BibitemShut {NoStop}%
\bibitem [{\citenamefont {Zhang}\ \emph {et~al.}(2023)\citenamefont {Zhang}, \citenamefont {Wang}, \citenamefont {Zhang}, \citenamefont {Rong}, \citenamefont {Cai}, \citenamefont {Wang}, \citenamefont {Ma}, \citenamefont {Guo}, \citenamefont {Chen}, \citenamefont {Wang} \emph {et~al.}}]{zhang2023multiple}%
  \BibitemOpen
  \bibfield  {author} {\bibinfo {author} {\bibfnamefont {C.}~\bibnamefont {Zhang}}, \bibinfo {author} {\bibfnamefont {Y.}~\bibnamefont {Wang}}, \bibinfo {author} {\bibfnamefont {F.}~\bibnamefont {Zhang}}, \bibinfo {author} {\bibfnamefont {H.}~\bibnamefont {Rong}}, \bibinfo {author} {\bibfnamefont {Y.}~\bibnamefont {Cai}}, \bibinfo {author} {\bibfnamefont {L.}~\bibnamefont {Wang}}, \bibinfo {author} {\bibfnamefont {X.-M.}\ \bibnamefont {Ma}}, \bibinfo {author} {\bibfnamefont {S.}~\bibnamefont {Guo}}, \bibinfo {author} {\bibfnamefont {Z.}~\bibnamefont {Chen}}, \bibinfo {author} {\bibfnamefont {Y.}~\bibnamefont {Wang}}, \emph {et~al.},\ }\bibfield  {title} {\bibinfo {title} {Multiple surface states, nontrivial band topology, and antiferromagnetism in \ce{GdAuAl4Ge2}},\ }\href@noop {} {\bibfield  {journal} {\bibinfo  {journal} {Chinese Physics B}\ }\textbf {\bibinfo {volume} {32}},\ \bibinfo {pages} {077401} (\bibinfo {year} {2023})}\BibitemShut {NoStop}%
\bibitem [{\citenamefont {Wu}\ and\ \citenamefont {Kanatzidis}(2005)}]{wu2005reaual4ge2}%
  \BibitemOpen
  \bibfield  {author} {\bibinfo {author} {\bibfnamefont {X.}~\bibnamefont {Wu}}\ and\ \bibinfo {author} {\bibfnamefont {M.~G.}\ \bibnamefont {Kanatzidis}},\ }\bibfield  {title} {\bibinfo {title} {{\ce{REAuAl4Ge2} and \ce{REAuAl4} \ce{(AuxGe_{1-x})2} (RE= rare earth element): Quaternary intermetallics grown in liquid aluminum}},\ }\href@noop {} {\bibfield  {journal} {\bibinfo  {journal} {Journal of Solid State Chemistry}\ }\textbf {\bibinfo {volume} {178}},\ \bibinfo {pages} {3233} (\bibinfo {year} {2005})}\BibitemShut {NoStop}%
\bibitem [{\citenamefont {Rezende-Gon{\c{c}}alves}\ \emph {et~al.}(2025)\citenamefont {Rezende-Gon{\c{c}}alves}, \citenamefont {Antezak}, \citenamefont {Kato}, \citenamefont {Feng}, \citenamefont {Fortuna}, \citenamefont {Le~F{\`e}vre}, \citenamefont {Rosmus}, \citenamefont {Olszowska}, \citenamefont {Sobol}, \citenamefont {Singh} \emph {et~al.}}]{rezende2025metallic}%
  \BibitemOpen
  \bibfield  {author} {\bibinfo {author} {\bibfnamefont {P.}~\bibnamefont {Rezende-Gon{\c{c}}alves}}, \bibinfo {author} {\bibfnamefont {A.}~\bibnamefont {Antezak}}, \bibinfo {author} {\bibfnamefont {T.}~\bibnamefont {Kato}}, \bibinfo {author} {\bibfnamefont {K.}~\bibnamefont {Feng}}, \bibinfo {author} {\bibfnamefont {F.}~\bibnamefont {Fortuna}}, \bibinfo {author} {\bibfnamefont {P.}~\bibnamefont {Le~F{\`e}vre}}, \bibinfo {author} {\bibfnamefont {M.}~\bibnamefont {Rosmus}}, \bibinfo {author} {\bibfnamefont {N.}~\bibnamefont {Olszowska}}, \bibinfo {author} {\bibfnamefont {T.}~\bibnamefont {Sobol}}, \bibinfo {author} {\bibfnamefont {D.}~\bibnamefont {Singh}}, \emph {et~al.},\ }\bibfield  {title} {\bibinfo {title} {{Metallic layered materials with magnetic frustration: An ARPES view of \ce{SmAuAl4Ge2} and \ce{TbAuAl4Ge2}}},\ }\href@noop {} {\bibfield  {journal} {\bibinfo  {journal} {Physical Review B}\ }\textbf {\bibinfo {volume} {111}},\ \bibinfo {pages} {205114} (\bibinfo {year} {2025})}\BibitemShut {NoStop}%
\bibitem [{\citenamefont {Feng}\ \emph {et~al.}(2024)\citenamefont {Feng}, \citenamefont {Bush}, \citenamefont {Oladehin}, \citenamefont {Lee},\ and\ \citenamefont {Baumbach}}]{feng2024complex}%
  \BibitemOpen
  \bibfield  {author} {\bibinfo {author} {\bibfnamefont {K.}~\bibnamefont {Feng}}, \bibinfo {author} {\bibfnamefont {C.}~\bibnamefont {Bush}}, \bibinfo {author} {\bibfnamefont {O.}~\bibnamefont {Oladehin}}, \bibinfo {author} {\bibfnamefont {M.}~\bibnamefont {Lee}},\ and\ \bibinfo {author} {\bibfnamefont {R.}~\bibnamefont {Baumbach}},\ }\bibfield  {title} {\bibinfo {title} {Complex antiferromagnetic order in the metallic triangular lattice compound \ce{SmAuAl4Ge2}},\ }\href@noop {} {\bibfield  {journal} {\bibinfo  {journal} {Physical Review B}\ }\textbf {\bibinfo {volume} {109}},\ \bibinfo {pages} {014436} (\bibinfo {year} {2024})}\BibitemShut {NoStop}%
\bibitem [{\citenamefont {Leahy}\ \emph {et~al.}(2022)\citenamefont {Leahy}, \citenamefont {Feng}, \citenamefont {Dery}, \citenamefont {Baumbach},\ and\ \citenamefont {Lee}}]{leahy2022field}%
  \BibitemOpen
  \bibfield  {author} {\bibinfo {author} {\bibfnamefont {I.~A.}\ \bibnamefont {Leahy}}, \bibinfo {author} {\bibfnamefont {K.}~\bibnamefont {Feng}}, \bibinfo {author} {\bibfnamefont {R.}~\bibnamefont {Dery}}, \bibinfo {author} {\bibfnamefont {R.}~\bibnamefont {Baumbach}},\ and\ \bibinfo {author} {\bibfnamefont {M.}~\bibnamefont {Lee}},\ }\bibfield  {title} {\bibinfo {title} {Field-induced magnetic states in the metallic rare-earth layered triangular antiferromagnet \ce{TbAuAl4Ge2}},\ }\href@noop {} {\bibfield  {journal} {\bibinfo  {journal} {Physical Review B}\ }\textbf {\bibinfo {volume} {106}},\ \bibinfo {pages} {094426} (\bibinfo {year} {2022})}\BibitemShut {NoStop}%
\bibitem [{\citenamefont {Feng}\ \emph {et~al.}(2023)\citenamefont {Feng}, \citenamefont {Cabral}, \citenamefont {Wei},\ and\ \citenamefont {Baumbach}}]{feng2023revealing}%
  \BibitemOpen
  \bibfield  {author} {\bibinfo {author} {\bibfnamefont {K.}~\bibnamefont {Feng}}, \bibinfo {author} {\bibfnamefont {J.~R.~G.}\ \bibnamefont {Cabral}}, \bibinfo {author} {\bibfnamefont {K.}~\bibnamefont {Wei}},\ and\ \bibinfo {author} {\bibfnamefont {R.}~\bibnamefont {Baumbach}},\ }\bibfield  {title} {\bibinfo {title} {Revealing complex spin states in \ce{GdNiAl4Ge2}},\ }\href@noop {} {\bibfield  {journal} {\bibinfo  {journal} {Physical Review Materials}\ }\textbf {\bibinfo {volume} {7}},\ \bibinfo {pages} {124409} (\bibinfo {year} {2023})}\BibitemShut {NoStop}%
\bibitem [{\citenamefont {Cong}\ \emph {et~al.}(2023)\citenamefont {Cong}, \citenamefont {Ge}, \citenamefont {Zhang}, \citenamefont {Ren}, \citenamefont {Zhao}, \citenamefont {Li}, \citenamefont {Wang}, \citenamefont {Zhu}, \citenamefont {Mei}, \citenamefont {Zhang} \emph {et~al.}}]{cong2023magnetic}%
  \BibitemOpen
  \bibfield  {author} {\bibinfo {author} {\bibfnamefont {M.}~\bibnamefont {Cong}}, \bibinfo {author} {\bibfnamefont {H.}~\bibnamefont {Ge}}, \bibinfo {author} {\bibfnamefont {L.}~\bibnamefont {Zhang}}, \bibinfo {author} {\bibfnamefont {W.}~\bibnamefont {Ren}}, \bibinfo {author} {\bibfnamefont {N.}~\bibnamefont {Zhao}}, \bibinfo {author} {\bibfnamefont {T.}~\bibnamefont {Li}}, \bibinfo {author} {\bibfnamefont {S.}~\bibnamefont {Wang}}, \bibinfo {author} {\bibfnamefont {J.}~\bibnamefont {Zhu}}, \bibinfo {author} {\bibfnamefont {J.}~\bibnamefont {Mei}}, \bibinfo {author} {\bibfnamefont {Q.}~\bibnamefont {Zhang}}, \emph {et~al.},\ }\bibfield  {title} {\bibinfo {title} {{Magnetic phase diagram and multiple field-induced states in the intermetallic triangular-lattice antiferromagnet \ce{NdAuAl4Ge2} with Ising-like spins}},\ }\href@noop {} {\bibfield  {journal} {\bibinfo  {journal} {Physical Review Materials}\ }\textbf {\bibinfo {volume} {7}},\ \bibinfo {pages} {024423} (\bibinfo {year} {2023})}\BibitemShut
  {NoStop}%
\bibitem [{\citenamefont {Wu}\ \emph {et~al.}(2025)\citenamefont {Wu}, \citenamefont {Ren}, \citenamefont {Calder}, \citenamefont {Zhang}, \citenamefont {Gao}, \citenamefont {An}, \citenamefont {Li},\ and\ \citenamefont {Zhang}}]{wu2025collinear}%
  \BibitemOpen
  \bibfield  {author} {\bibinfo {author} {\bibfnamefont {H.}~\bibnamefont {Wu}}, \bibinfo {author} {\bibfnamefont {W.}~\bibnamefont {Ren}}, \bibinfo {author} {\bibfnamefont {S.}~\bibnamefont {Calder}}, \bibinfo {author} {\bibfnamefont {Q.}~\bibnamefont {Zhang}}, \bibinfo {author} {\bibfnamefont {F.}~\bibnamefont {Gao}}, \bibinfo {author} {\bibfnamefont {M.}~\bibnamefont {An}}, \bibinfo {author} {\bibfnamefont {B.}~\bibnamefont {Li}},\ and\ \bibinfo {author} {\bibfnamefont {Z.}~\bibnamefont {Zhang}},\ }\bibfield  {title} {\bibinfo {title} {Collinear antiferromagnetic order in a quasi-two-dimensional triangular lattice compound \ce{DyNiAl4Ge2}},\ }\href@noop {} {\bibfield  {journal} {\bibinfo  {journal} {Physical Review Materials}\ }\textbf {\bibinfo {volume} {9}},\ \bibinfo {pages} {114402} (\bibinfo {year} {2025})}\BibitemShut {NoStop}%
\bibitem [{\citenamefont {Zhang}\ \emph {et~al.}(2017)\citenamefont {Zhang}, \citenamefont {Aryal}, \citenamefont {Huang}, \citenamefont {Chen}, \citenamefont {Lai}, \citenamefont {Graf}, \citenamefont {Besara}, \citenamefont {Siegrist}, \citenamefont {Manousakis},\ and\ \citenamefont {Baumbach}}]{zhang2017electronic}%
  \BibitemOpen
  \bibfield  {author} {\bibinfo {author} {\bibfnamefont {S.}~\bibnamefont {Zhang}}, \bibinfo {author} {\bibfnamefont {N.}~\bibnamefont {Aryal}}, \bibinfo {author} {\bibfnamefont {K.}~\bibnamefont {Huang}}, \bibinfo {author} {\bibfnamefont {K.-W.}\ \bibnamefont {Chen}}, \bibinfo {author} {\bibfnamefont {Y.}~\bibnamefont {Lai}}, \bibinfo {author} {\bibfnamefont {D.}~\bibnamefont {Graf}}, \bibinfo {author} {\bibfnamefont {T.}~\bibnamefont {Besara}}, \bibinfo {author} {\bibfnamefont {T.}~\bibnamefont {Siegrist}}, \bibinfo {author} {\bibfnamefont {E.}~\bibnamefont {Manousakis}},\ and\ \bibinfo {author} {\bibfnamefont {R.}~\bibnamefont {Baumbach}},\ }\bibfield  {title} {\bibinfo {title} {Electronic structure and magnetism in the layered triangular lattice compound \ce{CeAuAl4Ge2}},\ }\href@noop {} {\bibfield  {journal} {\bibinfo  {journal} {Physical Review Materials}\ }\textbf {\bibinfo {volume} {1}},\ \bibinfo {pages} {044404} (\bibinfo {year} {2017})}\BibitemShut {NoStop}%
\bibitem [{\citenamefont {Shin}\ \emph {et~al.}(2020)\citenamefont {Shin}, \citenamefont {Pomjakushin}, \citenamefont {Keller}, \citenamefont {Rosa}, \citenamefont {Stuhr}, \citenamefont {Niedermayer}, \citenamefont {Sibille}, \citenamefont {Toth}, \citenamefont {Kim}, \citenamefont {Jang} \emph {et~al.}}]{shin2020magnetic}%
  \BibitemOpen
  \bibfield  {author} {\bibinfo {author} {\bibfnamefont {S.}~\bibnamefont {Shin}}, \bibinfo {author} {\bibfnamefont {V.}~\bibnamefont {Pomjakushin}}, \bibinfo {author} {\bibfnamefont {L.}~\bibnamefont {Keller}}, \bibinfo {author} {\bibfnamefont {P.}~\bibnamefont {Rosa}}, \bibinfo {author} {\bibfnamefont {U.}~\bibnamefont {Stuhr}}, \bibinfo {author} {\bibfnamefont {C.}~\bibnamefont {Niedermayer}}, \bibinfo {author} {\bibfnamefont {R.}~\bibnamefont {Sibille}}, \bibinfo {author} {\bibfnamefont {S.}~\bibnamefont {Toth}}, \bibinfo {author} {\bibfnamefont {J.}~\bibnamefont {Kim}}, \bibinfo {author} {\bibfnamefont {H.}~\bibnamefont {Jang}}, \emph {et~al.},\ }\bibfield  {title} {\bibinfo {title} {Magnetic structure and crystalline electric field effects in the triangular antiferromagnet \ce{CePtAl4Ge2}},\ }\href@noop {} {\bibfield  {journal} {\bibinfo  {journal} {Physical Review B}\ }\textbf {\bibinfo {volume} {101}},\ \bibinfo {pages} {224421} (\bibinfo {year} {2020})}\BibitemShut {NoStop}%
\bibitem [{\citenamefont {Gao}\ \emph {et~al.}(2024)\citenamefont {Gao}, \citenamefont {Ren}, \citenamefont {Wang}, \citenamefont {Yano}, \citenamefont {Calder}, \citenamefont {Zhang}, \citenamefont {Wu}, \citenamefont {An}, \citenamefont {Jing}, \citenamefont {Li} \emph {et~al.}}]{gao2024high}%
  \BibitemOpen
  \bibfield  {author} {\bibinfo {author} {\bibfnamefont {F.}~\bibnamefont {Gao}}, \bibinfo {author} {\bibfnamefont {W.}~\bibnamefont {Ren}}, \bibinfo {author} {\bibfnamefont {C.-W.}\ \bibnamefont {Wang}}, \bibinfo {author} {\bibfnamefont {S.}~\bibnamefont {Yano}}, \bibinfo {author} {\bibfnamefont {S.}~\bibnamefont {Calder}}, \bibinfo {author} {\bibfnamefont {Q.}~\bibnamefont {Zhang}}, \bibinfo {author} {\bibfnamefont {H.}~\bibnamefont {Wu}}, \bibinfo {author} {\bibfnamefont {M.}~\bibnamefont {An}}, \bibinfo {author} {\bibfnamefont {Y.}~\bibnamefont {Jing}}, \bibinfo {author} {\bibfnamefont {B.}~\bibnamefont {Li}}, \emph {et~al.},\ }\bibfield  {title} {\bibinfo {title} {High-order harmonics and the reverse of the squaring up process in the triangular-lattice magnet \ce{HoPdAl4Ge2}},\ }\href@noop {} {\bibfield  {journal} {\bibinfo  {journal} {Physical Review B}\ }\textbf {\bibinfo {volume} {109}},\ \bibinfo {pages} {134407} (\bibinfo {year} {2024})}\BibitemShut {NoStop}%
\bibitem [{\citenamefont {Sieve}\ \emph {et~al.}(2002)\citenamefont {Sieve}, \citenamefont {Trikalitis},\ and\ \citenamefont {Kanatzidis}}]{sieve2002quaternary}%
  \BibitemOpen
  \bibfield  {author} {\bibinfo {author} {\bibfnamefont {B.}~\bibnamefont {Sieve}}, \bibinfo {author} {\bibfnamefont {P.~N.}\ \bibnamefont {Trikalitis}},\ and\ \bibinfo {author} {\bibfnamefont {M.~G.}\ \bibnamefont {Kanatzidis}},\ }\bibfield  {title} {\bibinfo {title} {{Quaternary Germanides Formed in Molten Aluminum: \ce{Tb2NiAl4Ge2} and \ce{Ce2NiAl_{6-x}Ge_{4-y}} (x~ 0.24, y~ 1.34)}},\ }\href@noop {} {\bibfield  {journal} {\bibinfo  {journal} {Zeitschrift f{\"u}r anorganische und allgemeine Chemie}\ }\textbf {\bibinfo {volume} {628}},\ \bibinfo {pages} {1568} (\bibinfo {year} {2002})}\BibitemShut {NoStop}%
\bibitem [{\citenamefont {Demchenko}\ \emph {et~al.}(2008)\citenamefont {Demchenko}, \citenamefont {Ko{\'n}czyk}, \citenamefont {Demchenko}, \citenamefont {Gladyshevskii}, \citenamefont {Majzner},\ and\ \citenamefont {Muratova}}]{demchenko2008quaternary}%
  \BibitemOpen
  \bibfield  {author} {\bibinfo {author} {\bibfnamefont {G.}~\bibnamefont {Demchenko}}, \bibinfo {author} {\bibfnamefont {J.}~\bibnamefont {Ko{\'n}czyk}}, \bibinfo {author} {\bibfnamefont {P.}~\bibnamefont {Demchenko}}, \bibinfo {author} {\bibfnamefont {R.}~\bibnamefont {Gladyshevskii}}, \bibinfo {author} {\bibfnamefont {W.}~\bibnamefont {Majzner}},\ and\ \bibinfo {author} {\bibfnamefont {L.}~\bibnamefont {Muratova}},\ }\bibfield  {title} {\bibinfo {title} {{Quaternary alumogermanides in the Er}},\ }\href@noop {} {\bibfield  {journal} {\bibinfo  {journal} {Chemistry of metals and alloys}\ } (\bibinfo {year} {2008})}\BibitemShut {NoStop}%
\bibitem [{\citenamefont {Semusyo}\ \emph {et~al.}(2015)\citenamefont {Semusyo}, \citenamefont {Lutsishyn}, \citenamefont {Pukas}, \citenamefont {Tokaychuk},\ and\ \citenamefont {Gladyshevsky}}]{semuso2015tokaychuk}%
  \BibitemOpen
  \bibfield  {author} {\bibinfo {author} {\bibfnamefont {N.~Z.}\ \bibnamefont {Semusyo}}, \bibinfo {author} {\bibfnamefont {Y.~Y.}\ \bibnamefont {Lutsishyn}}, \bibinfo {author} {\bibfnamefont {S.~Y.}\ \bibnamefont {Pukas}}, \bibinfo {author} {\bibfnamefont {Y.~O.}\ \bibnamefont {Tokaychuk}},\ and\ \bibinfo {author} {\bibfnamefont {R.}~\bibnamefont {Gladyshevsky}},\ }\bibfield  {title} {\bibinfo {title} {New tetrahedral aluminum germanides with a structure of the type \ce{Tb2NiAl4Ge2}},\ }\href@noop {} {\bibfield  {journal} {\bibinfo  {journal} {Ukrainian Chemical Journal}\ }\textbf {\bibinfo {volume} {81}},\ \bibinfo {pages} {36} (\bibinfo {year} {2015})}\BibitemShut {NoStop}%
\bibitem [{\citenamefont {He}\ \emph {et~al.}(2015)\citenamefont {He}, \citenamefont {Zeng}, \citenamefont {Yang},\ and\ \citenamefont {Lin}}]{he2015crystal}%
  \BibitemOpen
  \bibfield  {author} {\bibinfo {author} {\bibfnamefont {W.}~\bibnamefont {He}}, \bibinfo {author} {\bibfnamefont {W.}~\bibnamefont {Zeng}}, \bibinfo {author} {\bibfnamefont {T.}~\bibnamefont {Yang}},\ and\ \bibinfo {author} {\bibfnamefont {G.}~\bibnamefont {Lin}},\ }\bibfield  {title} {\bibinfo {title} {{Crystal structure of new \ce{R2TAl4Ge2} (R= Y, Gd--Er, T= Fe, Co) quaternary compounds and magnetic properties of \ce{Gd2TAl4Ge2}}},\ }\href@noop {} {\bibfield  {journal} {\bibinfo  {journal} {Journal of Alloys and Compounds}\ }\textbf {\bibinfo {volume} {633}},\ \bibinfo {pages} {265} (\bibinfo {year} {2015})}\BibitemShut {NoStop}%
\bibitem [{\citenamefont {Pukas}\ and\ \citenamefont {Gladyshevskii}(2024)}]{pukas2024quaternary}%
  \BibitemOpen
  \bibfield  {author} {\bibinfo {author} {\bibfnamefont {S.}~\bibnamefont {Pukas}}\ and\ \bibinfo {author} {\bibfnamefont {R.}~\bibnamefont {Gladyshevskii}},\ }\bibfield  {title} {\bibinfo {title} {{Quaternary compounds in R--T--Al--M systems}},\ }in\ \href@noop {} {\emph {\bibinfo {booktitle} {ACTA CRYSTALLOGRAPHICA A-FOUNDATION AND ADVANCES}}},\ Vol.~\bibinfo {volume} {80}\ (\bibinfo {organization} {INT UNION CRYSTALLOGRAPHY 2 ABBEY SQ, CHESTER, CH1 2HU, ENGLAND},\ \bibinfo {year} {2024})\BibitemShut {NoStop}%
\bibitem [{\citenamefont {Huang}\ \emph {et~al.}(2019)\citenamefont {Huang}, \citenamefont {Sun}, \citenamefont {Sun}, \citenamefont {Zhang},\ and\ \citenamefont {Lei}}]{huang2019physical}%
  \BibitemOpen
  \bibfield  {author} {\bibinfo {author} {\bibfnamefont {K.}~\bibnamefont {Huang}}, \bibinfo {author} {\bibfnamefont {Y.}~\bibnamefont {Sun}}, \bibinfo {author} {\bibfnamefont {S.}~\bibnamefont {Sun}}, \bibinfo {author} {\bibfnamefont {X.}~\bibnamefont {Zhang}},\ and\ \bibinfo {author} {\bibfnamefont {H.}~\bibnamefont {Lei}},\ }\bibfield  {title} {\bibinfo {title} {{Physical properties of quaternary compounds \ce{Gd2CoAl4T2} (T= Si, Ge) single crystals}},\ }\href@noop {} {\bibfield  {journal} {\bibinfo  {journal} {Frontiers of Physics}\ }\textbf {\bibinfo {volume} {14}},\ \bibinfo {pages} {23502} (\bibinfo {year} {2019})}\BibitemShut {NoStop}%
\bibitem [{\citenamefont {Canfield}\ \emph {et~al.}(2016)\citenamefont {Canfield}, \citenamefont {Kong}, \citenamefont {Kaluarachchi},\ and\ \citenamefont {Jo}}]{canfield2016use}%
  \BibitemOpen
  \bibfield  {author} {\bibinfo {author} {\bibfnamefont {P.~C.}\ \bibnamefont {Canfield}}, \bibinfo {author} {\bibfnamefont {T.}~\bibnamefont {Kong}}, \bibinfo {author} {\bibfnamefont {U.~S.}\ \bibnamefont {Kaluarachchi}},\ and\ \bibinfo {author} {\bibfnamefont {N.~H.}\ \bibnamefont {Jo}},\ }\bibfield  {title} {\bibinfo {title} {Use of frit-disc crucibles for routine and exploratory solution growth of single crystalline samples},\ }\href@noop {} {\bibfield  {journal} {\bibinfo  {journal} {Philosophical Magazine}\ }\textbf {\bibinfo {volume} {96}},\ \bibinfo {pages} {84} (\bibinfo {year} {2016})}\BibitemShut {NoStop}%
\bibitem [{\citenamefont {Latturner}\ \emph {et~al.}(2002)\citenamefont {Latturner}, \citenamefont {Bilc}, \citenamefont {Mahanti},\ and\ \citenamefont {Kanatzidis}}]{latturner2002quaternary}%
  \BibitemOpen
  \bibfield  {author} {\bibinfo {author} {\bibfnamefont {S.}~\bibnamefont {Latturner}}, \bibinfo {author} {\bibfnamefont {D.}~\bibnamefont {Bilc}}, \bibinfo {author} {\bibfnamefont {S.}~\bibnamefont {Mahanti}},\ and\ \bibinfo {author} {\bibfnamefont {M.}~\bibnamefont {Kanatzidis}},\ }\bibfield  {title} {\bibinfo {title} {{Quaternary Intermetallics Grown from Molten Aluminum: The Homologous Series \ce{Th2(Au_{x}Si_{1-x})[AuAl2]_{n}Si2} (n= 1, 2, 4)}},\ }\href@noop {} {\bibfield  {journal} {\bibinfo  {journal} {Chemistry of materials}\ }\textbf {\bibinfo {volume} {14}},\ \bibinfo {pages} {1695} (\bibinfo {year} {2002})}\BibitemShut {NoStop}%
\bibitem [{Sup()}]{Supplemental}%
  \BibitemOpen
  \href@noop {} {\bibinfo {title} {{See Supplemental Material for Crystallographic Information Files.}}},\ \bibinfo {howpublished} {\url{provided by publisher}}\BibitemShut {NoStop}%
\bibitem [{\citenamefont {Parthe}\ \emph {et~al.}(1993)\citenamefont {Parthe}, \citenamefont {Gelato}, \citenamefont {Chabot}, \citenamefont {Penzo}, \citenamefont {Cenzuel},\ and\ \citenamefont {Gladyshevskii}}]{parthe1993table}%
  \BibitemOpen
  \bibfield  {author} {\bibinfo {author} {\bibfnamefont {E.}~\bibnamefont {Parthe}}, \bibinfo {author} {\bibfnamefont {L.}~\bibnamefont {Gelato}}, \bibinfo {author} {\bibfnamefont {B.}~\bibnamefont {Chabot}}, \bibinfo {author} {\bibfnamefont {M.}~\bibnamefont {Penzo}}, \bibinfo {author} {\bibfnamefont {K.}~\bibnamefont {Cenzuel}},\ and\ \bibinfo {author} {\bibfnamefont {R.}~\bibnamefont {Gladyshevskii}},\ }\bibfield  {title} {\bibinfo {title} {{Table E22 in Volume 1 of ‘‘TYPIX, Standard Data and Crystal Chemical Characterization of Inorganic Structure Types,’’}},\ }\href@noop {} {\bibfield  {journal} {\bibinfo  {journal} {Gmelin Handbook of Inorganic and Organometallic Chemistry, Berlin: Springer}\ } (\bibinfo {year} {1993})}\BibitemShut {NoStop}%
\bibitem [{\citenamefont {Bruinsma}\ and\ \citenamefont {Aeppli}(1984)}]{bruinsma1984metamagnets}%
  \BibitemOpen
  \bibfield  {author} {\bibinfo {author} {\bibfnamefont {R.}~\bibnamefont {Bruinsma}}\ and\ \bibinfo {author} {\bibfnamefont {G.}~\bibnamefont {Aeppli}},\ }\bibfield  {title} {\bibinfo {title} {Metamagnets and frustration},\ }\href@noop {} {\bibfield  {journal} {\bibinfo  {journal} {Physical Review B}\ }\textbf {\bibinfo {volume} {29}},\ \bibinfo {pages} {2644} (\bibinfo {year} {1984})}\BibitemShut {NoStop}%
\bibitem [{\citenamefont {Rau}\ \emph {et~al.}(2016)\citenamefont {Rau}, \citenamefont {Lee},\ and\ \citenamefont {Kee}}]{rau2016spin}%
  \BibitemOpen
  \bibfield  {author} {\bibinfo {author} {\bibfnamefont {J.~G.}\ \bibnamefont {Rau}}, \bibinfo {author} {\bibfnamefont {E.~K.-H.}\ \bibnamefont {Lee}},\ and\ \bibinfo {author} {\bibfnamefont {H.-Y.}\ \bibnamefont {Kee}},\ }\bibfield  {title} {\bibinfo {title} {{Spin-orbit physics giving rise to novel phases in correlated systems: Iridates and related materials}},\ }\href@noop {} {\bibfield  {journal} {\bibinfo  {journal} {Annual Review of Condensed Matter Physics}\ }\textbf {\bibinfo {volume} {7}},\ \bibinfo {pages} {195} (\bibinfo {year} {2016})}\BibitemShut {NoStop}%
\bibitem [{\citenamefont {Schaffer}\ \emph {et~al.}(2016)\citenamefont {Schaffer}, \citenamefont {Kin-Ho~Lee}, \citenamefont {Yang},\ and\ \citenamefont {Kim}}]{schaffer2016recent}%
  \BibitemOpen
  \bibfield  {author} {\bibinfo {author} {\bibfnamefont {R.}~\bibnamefont {Schaffer}}, \bibinfo {author} {\bibfnamefont {E.}~\bibnamefont {Kin-Ho~Lee}}, \bibinfo {author} {\bibfnamefont {B.-J.}\ \bibnamefont {Yang}},\ and\ \bibinfo {author} {\bibfnamefont {Y.~B.}\ \bibnamefont {Kim}},\ }\bibfield  {title} {\bibinfo {title} {Recent progress on correlated electron systems with strong spin--orbit coupling},\ }\href@noop {} {\bibfield  {journal} {\bibinfo  {journal} {Reports on Progress in Physics}\ }\textbf {\bibinfo {volume} {79}},\ \bibinfo {pages} {094504} (\bibinfo {year} {2016})}\BibitemShut {NoStop}%
\bibitem [{\citenamefont {Li}\ \emph {et~al.}(2022)\citenamefont {Li}, \citenamefont {Huang}, \citenamefont {Yue}, \citenamefont {Guang}, \citenamefont {Xia}, \citenamefont {Wang}, \citenamefont {Li}, \citenamefont {Zhao}, \citenamefont {Zhou},\ and\ \citenamefont {Sun}}]{li2022low}%
  \BibitemOpen
  \bibfield  {author} {\bibinfo {author} {\bibfnamefont {N.}~\bibnamefont {Li}}, \bibinfo {author} {\bibfnamefont {Q.}~\bibnamefont {Huang}}, \bibinfo {author} {\bibfnamefont {X.}~\bibnamefont {Yue}}, \bibinfo {author} {\bibfnamefont {S.}~\bibnamefont {Guang}}, \bibinfo {author} {\bibfnamefont {K.}~\bibnamefont {Xia}}, \bibinfo {author} {\bibfnamefont {Y.}~\bibnamefont {Wang}}, \bibinfo {author} {\bibfnamefont {Q.}~\bibnamefont {Li}}, \bibinfo {author} {\bibfnamefont {X.}~\bibnamefont {Zhao}}, \bibinfo {author} {\bibfnamefont {H.}~\bibnamefont {Zhou}},\ and\ \bibinfo {author} {\bibfnamefont {X.}~\bibnamefont {Sun}},\ }\bibfield  {title} {\bibinfo {title} {{Low-temperature transport properties of the intermetallic compound HoAgGe with a kagome spin-ice state}},\ }\href@noop {} {\bibfield  {journal} {\bibinfo  {journal} {Physical Review B}\ }\textbf {\bibinfo {volume} {106}},\ \bibinfo {pages} {014416} (\bibinfo {year} {2022})}\BibitemShut {NoStop}%
\bibitem [{\citenamefont {Zhao}\ \emph {et~al.}(2020)\citenamefont {Zhao}, \citenamefont {Deng}, \citenamefont {Chen}, \citenamefont {Ross}, \citenamefont {Pet{\v{r}}{\'\i}{\v{c}}ek}, \citenamefont {G{\"u}nther}, \citenamefont {Russina}, \citenamefont {Hutanu},\ and\ \citenamefont {Gegenwart}}]{zhao2020realization}%
  \BibitemOpen
  \bibfield  {author} {\bibinfo {author} {\bibfnamefont {K.}~\bibnamefont {Zhao}}, \bibinfo {author} {\bibfnamefont {H.}~\bibnamefont {Deng}}, \bibinfo {author} {\bibfnamefont {H.}~\bibnamefont {Chen}}, \bibinfo {author} {\bibfnamefont {K.~A.}\ \bibnamefont {Ross}}, \bibinfo {author} {\bibfnamefont {V.}~\bibnamefont {Pet{\v{r}}{\'\i}{\v{c}}ek}}, \bibinfo {author} {\bibfnamefont {G.}~\bibnamefont {G{\"u}nther}}, \bibinfo {author} {\bibfnamefont {M.}~\bibnamefont {Russina}}, \bibinfo {author} {\bibfnamefont {V.}~\bibnamefont {Hutanu}},\ and\ \bibinfo {author} {\bibfnamefont {P.}~\bibnamefont {Gegenwart}},\ }\bibfield  {title} {\bibinfo {title} {Realization of the kagome spin ice state in a frustrated intermetallic compound},\ }\href@noop {} {\bibfield  {journal} {\bibinfo  {journal} {Science}\ }\textbf {\bibinfo {volume} {367}},\ \bibinfo {pages} {1218} (\bibinfo {year} {2020})}\BibitemShut {NoStop}%
\bibitem [{\citenamefont {Kurumaji}\ \emph {et~al.}(2024)\citenamefont {Kurumaji}, \citenamefont {Fang}, \citenamefont {Ye}, \citenamefont {Kitou},\ and\ \citenamefont {Checkelsky}}]{kurumaji2024metamagnetic}%
  \BibitemOpen
  \bibfield  {author} {\bibinfo {author} {\bibfnamefont {T.}~\bibnamefont {Kurumaji}}, \bibinfo {author} {\bibfnamefont {S.}~\bibnamefont {Fang}}, \bibinfo {author} {\bibfnamefont {L.}~\bibnamefont {Ye}}, \bibinfo {author} {\bibfnamefont {S.}~\bibnamefont {Kitou}},\ and\ \bibinfo {author} {\bibfnamefont {J.~G.}\ \bibnamefont {Checkelsky}},\ }\bibfield  {title} {\bibinfo {title} {{Metamagnetic multiband Hall effect in Ising antiferromagnet \ce{ErGa2}}},\ }\href@noop {} {\bibfield  {journal} {\bibinfo  {journal} {Proceedings of the National Academy of Sciences}\ }\textbf {\bibinfo {volume} {121}},\ \bibinfo {pages} {e2318411121} (\bibinfo {year} {2024})}\BibitemShut {NoStop}%
\bibitem [{\citenamefont {Doukour{\'e}}\ and\ \citenamefont {Gignoux}(1982)}]{doukoure1982metamagnetism}%
  \BibitemOpen
  \bibfield  {author} {\bibinfo {author} {\bibfnamefont {M.}~\bibnamefont {Doukour{\'e}}}\ and\ \bibinfo {author} {\bibfnamefont {D.}~\bibnamefont {Gignoux}},\ }\bibfield  {title} {\bibinfo {title} {Metamagnetism in \ce{ErGa2} studied on a single crystal},\ }\href@noop {} {\bibfield  {journal} {\bibinfo  {journal} {Journal of magnetism and magnetic materials}\ }\textbf {\bibinfo {volume} {30}},\ \bibinfo {pages} {111} (\bibinfo {year} {1982})}\BibitemShut {NoStop}%
\bibitem [{\citenamefont {Kusakabe}\ \emph {et~al.}(2025)\citenamefont {Kusakabe}, \citenamefont {Mikawa}, \citenamefont {Takei}, \citenamefont {Katsufuji}, \citenamefont {Kawasugi},\ and\ \citenamefont {Suzuki}}]{kusakabe2025magnetotransport}%
  \BibitemOpen
  \bibfield  {author} {\bibinfo {author} {\bibfnamefont {S.}~\bibnamefont {Kusakabe}}, \bibinfo {author} {\bibfnamefont {R.}~\bibnamefont {Mikawa}}, \bibinfo {author} {\bibfnamefont {H.}~\bibnamefont {Takei}}, \bibinfo {author} {\bibfnamefont {T.}~\bibnamefont {Katsufuji}}, \bibinfo {author} {\bibfnamefont {Y.}~\bibnamefont {Kawasugi}},\ and\ \bibinfo {author} {\bibfnamefont {T.}~\bibnamefont {Suzuki}},\ }\bibfield  {title} {\bibinfo {title} {Magnetotransport of the square-net topological semimetal with magnetization plateaus \ce{DyAgSb2}},\ }\href@noop {} {\bibfield  {journal} {\bibinfo  {journal} {Physical Review B}\ }\textbf {\bibinfo {volume} {112}},\ \bibinfo {pages} {205119} (\bibinfo {year} {2025})}\BibitemShut {NoStop}%
\bibitem [{\citenamefont {Guo}\ \emph {et~al.}(2024)\citenamefont {Guo}, \citenamefont {Ma}, \citenamefont {Liu}, \citenamefont {Wu}, \citenamefont {Wang}, \citenamefont {Shi}, \citenamefont {Li},\ and\ \citenamefont {Jia}}]{guo20241}%
  \BibitemOpen
  \bibfield  {author} {\bibinfo {author} {\bibfnamefont {K.}~\bibnamefont {Guo}}, \bibinfo {author} {\bibfnamefont {Z.}~\bibnamefont {Ma}}, \bibinfo {author} {\bibfnamefont {H.}~\bibnamefont {Liu}}, \bibinfo {author} {\bibfnamefont {Z.}~\bibnamefont {Wu}}, \bibinfo {author} {\bibfnamefont {J.}~\bibnamefont {Wang}}, \bibinfo {author} {\bibfnamefont {Y.}~\bibnamefont {Shi}}, \bibinfo {author} {\bibfnamefont {Y.}~\bibnamefont {Li}},\ and\ \bibinfo {author} {\bibfnamefont {S.}~\bibnamefont {Jia}},\ }\bibfield  {title} {\bibinfo {title} {{1/3 and other magnetization plateaus in the quasi-one-dimensional Ising magnet \ce{TbTi3Bi4} with zigzag spin chain}},\ }\href@noop {} {\bibfield  {journal} {\bibinfo  {journal} {Physical Review B}\ }\textbf {\bibinfo {volume} {110}},\ \bibinfo {pages} {064416} (\bibinfo {year} {2024})}\BibitemShut {NoStop}%
\bibitem [{\citenamefont {Ortiz}\ \emph {et~al.}(2024)\citenamefont {Ortiz}, \citenamefont {Zhang}, \citenamefont {G{\'o}rnicka}, \citenamefont {Parker}, \citenamefont {Samolyuk}, \citenamefont {Yang}, \citenamefont {Miao}, \citenamefont {Lu}, \citenamefont {Moore}, \citenamefont {May} \emph {et~al.}}]{ortiz2024intricate}%
  \BibitemOpen
  \bibfield  {author} {\bibinfo {author} {\bibfnamefont {B.~R.}\ \bibnamefont {Ortiz}}, \bibinfo {author} {\bibfnamefont {H.}~\bibnamefont {Zhang}}, \bibinfo {author} {\bibfnamefont {K.}~\bibnamefont {G{\'o}rnicka}}, \bibinfo {author} {\bibfnamefont {D.~S.}\ \bibnamefont {Parker}}, \bibinfo {author} {\bibfnamefont {G.~D.}\ \bibnamefont {Samolyuk}}, \bibinfo {author} {\bibfnamefont {F.}~\bibnamefont {Yang}}, \bibinfo {author} {\bibfnamefont {H.}~\bibnamefont {Miao}}, \bibinfo {author} {\bibfnamefont {Q.}~\bibnamefont {Lu}}, \bibinfo {author} {\bibfnamefont {R.~G.}\ \bibnamefont {Moore}}, \bibinfo {author} {\bibfnamefont {A.~F.}\ \bibnamefont {May}}, \emph {et~al.},\ }\bibfield  {title} {\bibinfo {title} {{Intricate magnetic landscape in antiferromagnetic kagome metal \ce{TbTi3Bi4} and interplay with \ce{Ln$_{2-x}$Ti$_{6+x}$Bi9}(Ln: Tb{\textperiodcentered}{\textperiodcentered}{\textperiodcentered} Lu) shurikagome metals}},\ }\href@noop {} {\bibfield  {journal} {\bibinfo  {journal} {Chemistry of Materials}\
  }\textbf {\bibinfo {volume} {36}},\ \bibinfo {pages} {8002} (\bibinfo {year} {2024})}\BibitemShut {NoStop}%
\bibitem [{\citenamefont {Johnston}(2017)}]{johnston2017influence}%
  \BibitemOpen
  \bibfield  {author} {\bibinfo {author} {\bibfnamefont {D.~C.}\ \bibnamefont {Johnston}},\ }\bibfield  {title} {\bibinfo {title} {{Influence of classical anisotropy fields on the properties of Heisenberg antiferromagnets within unified molecular field theory}},\ }\href@noop {} {\bibfield  {journal} {\bibinfo  {journal} {Physical Review B}\ }\textbf {\bibinfo {volume} {96}},\ \bibinfo {pages} {224428} (\bibinfo {year} {2017})}\BibitemShut {NoStop}%
\bibitem [{\citenamefont {Du}\ \emph {et~al.}(2021)\citenamefont {Du}, \citenamefont {Lu},\ and\ \citenamefont {Xie}}]{du2021nonlinear}%
  \BibitemOpen
  \bibfield  {author} {\bibinfo {author} {\bibfnamefont {Z.}~\bibnamefont {Du}}, \bibinfo {author} {\bibfnamefont {H.-Z.}\ \bibnamefont {Lu}},\ and\ \bibinfo {author} {\bibfnamefont {X.}~\bibnamefont {Xie}},\ }\bibfield  {title} {\bibinfo {title} {Nonlinear hall effects},\ }\href@noop {} {\bibfield  {journal} {\bibinfo  {journal} {Nature Reviews Physics}\ }\textbf {\bibinfo {volume} {3}},\ \bibinfo {pages} {744} (\bibinfo {year} {2021})}\BibitemShut {NoStop}%
\bibitem [{\citenamefont {Nagaosa}\ \emph {et~al.}(2010)\citenamefont {Nagaosa}, \citenamefont {Sinova}, \citenamefont {Onoda}, \citenamefont {MacDonald},\ and\ \citenamefont {Ong}}]{nagaosa2010anomalous}%
  \BibitemOpen
  \bibfield  {author} {\bibinfo {author} {\bibfnamefont {N.}~\bibnamefont {Nagaosa}}, \bibinfo {author} {\bibfnamefont {J.}~\bibnamefont {Sinova}}, \bibinfo {author} {\bibfnamefont {S.}~\bibnamefont {Onoda}}, \bibinfo {author} {\bibfnamefont {A.~H.}\ \bibnamefont {MacDonald}},\ and\ \bibinfo {author} {\bibfnamefont {N.~P.}\ \bibnamefont {Ong}},\ }\bibfield  {title} {\bibinfo {title} {Anomalous hall effect},\ }\href@noop {} {\bibfield  {journal} {\bibinfo  {journal} {Reviews of modern physics}\ }\textbf {\bibinfo {volume} {82}},\ \bibinfo {pages} {1539} (\bibinfo {year} {2010})}\BibitemShut {NoStop}%
\bibitem [{\citenamefont {Zeng}\ \emph {et~al.}(2006)\citenamefont {Zeng}, \citenamefont {Yao}, \citenamefont {Niu},\ and\ \citenamefont {Weitering}}]{zeng2006linear}%
  \BibitemOpen
  \bibfield  {author} {\bibinfo {author} {\bibfnamefont {C.}~\bibnamefont {Zeng}}, \bibinfo {author} {\bibfnamefont {Y.}~\bibnamefont {Yao}}, \bibinfo {author} {\bibfnamefont {Q.}~\bibnamefont {Niu}},\ and\ \bibinfo {author} {\bibfnamefont {H.~H.}\ \bibnamefont {Weitering}},\ }\bibfield  {title} {\bibinfo {title} {{Linear magnetization dependence of the intrinsic anomalous Hall effect}},\ }\href@noop {} {\bibfield  {journal} {\bibinfo  {journal} {Physical review letters}\ }\textbf {\bibinfo {volume} {96}},\ \bibinfo {pages} {037204} (\bibinfo {year} {2006})}\BibitemShut {NoStop}%
\bibitem [{\citenamefont {Liu}\ \emph {et~al.}(2023)\citenamefont {Liu}, \citenamefont {Feng}, \citenamefont {Zhang}, \citenamefont {Deng}, \citenamefont {Dong}, \citenamefont {Zhang}, \citenamefont {Li}, \citenamefont {Lu}, \citenamefont {Chang},\ and\ \citenamefont {Wang}}]{liu2023topological}%
  \BibitemOpen
  \bibfield  {author} {\bibinfo {author} {\bibfnamefont {X.}~\bibnamefont {Liu}}, \bibinfo {author} {\bibfnamefont {Q.}~\bibnamefont {Feng}}, \bibinfo {author} {\bibfnamefont {D.}~\bibnamefont {Zhang}}, \bibinfo {author} {\bibfnamefont {Y.}~\bibnamefont {Deng}}, \bibinfo {author} {\bibfnamefont {S.}~\bibnamefont {Dong}}, \bibinfo {author} {\bibfnamefont {E.}~\bibnamefont {Zhang}}, \bibinfo {author} {\bibfnamefont {W.}~\bibnamefont {Li}}, \bibinfo {author} {\bibfnamefont {Q.}~\bibnamefont {Lu}}, \bibinfo {author} {\bibfnamefont {K.}~\bibnamefont {Chang}},\ and\ \bibinfo {author} {\bibfnamefont {K.}~\bibnamefont {Wang}},\ }\bibfield  {title} {\bibinfo {title} {Topological spin textures in a non-collinear antiferromagnet system},\ }\href@noop {} {\bibfield  {journal} {\bibinfo  {journal} {Advanced Materials}\ }\textbf {\bibinfo {volume} {35}},\ \bibinfo {pages} {2211634} (\bibinfo {year} {2023})}\BibitemShut {NoStop}%
\bibitem [{\citenamefont {Wang}\ \emph {et~al.}(2021)\citenamefont {Wang}, \citenamefont {Neubauer}, \citenamefont {Duan}, \citenamefont {Yin}, \citenamefont {Fujitsu}, \citenamefont {Hosono}, \citenamefont {Ye}, \citenamefont {Zhang}, \citenamefont {Chi}, \citenamefont {Krycka} \emph {et~al.}}]{wang2021field}%
  \BibitemOpen
  \bibfield  {author} {\bibinfo {author} {\bibfnamefont {Q.}~\bibnamefont {Wang}}, \bibinfo {author} {\bibfnamefont {K.~J.}\ \bibnamefont {Neubauer}}, \bibinfo {author} {\bibfnamefont {C.}~\bibnamefont {Duan}}, \bibinfo {author} {\bibfnamefont {Q.}~\bibnamefont {Yin}}, \bibinfo {author} {\bibfnamefont {S.}~\bibnamefont {Fujitsu}}, \bibinfo {author} {\bibfnamefont {H.}~\bibnamefont {Hosono}}, \bibinfo {author} {\bibfnamefont {F.}~\bibnamefont {Ye}}, \bibinfo {author} {\bibfnamefont {R.}~\bibnamefont {Zhang}}, \bibinfo {author} {\bibfnamefont {S.}~\bibnamefont {Chi}}, \bibinfo {author} {\bibfnamefont {K.}~\bibnamefont {Krycka}}, \emph {et~al.},\ }\bibfield  {title} {\bibinfo {title} {{Field-induced topological Hall effect and double-fan spin structure with ac-axis component in the metallic kagome antiferromagnetic compound \ce{YMn6Sn6}}},\ }\href@noop {} {\bibfield  {journal} {\bibinfo  {journal} {Physical Review B}\ }\textbf {\bibinfo {volume} {103}},\ \bibinfo {pages} {014416} (\bibinfo {year}
  {2021})}\BibitemShut {NoStop}%
\bibitem [{\citenamefont {Wang}\ \emph {et~al.}(2023)\citenamefont {Wang}, \citenamefont {Ma}, \citenamefont {Zeng}, \citenamefont {Gong}, \citenamefont {Lin}, \citenamefont {Wang}, \citenamefont {Dai}, \citenamefont {Han}, \citenamefont {Wang},\ and\ \citenamefont {Xia}}]{wang2023anomalous}%
  \BibitemOpen
  \bibfield  {author} {\bibinfo {author} {\bibfnamefont {H.}~\bibnamefont {Wang}}, \bibinfo {author} {\bibfnamefont {X.-P.}\ \bibnamefont {Ma}}, \bibinfo {author} {\bibfnamefont {X.-Y.}\ \bibnamefont {Zeng}}, \bibinfo {author} {\bibfnamefont {J.}~\bibnamefont {Gong}}, \bibinfo {author} {\bibfnamefont {J.-F.}\ \bibnamefont {Lin}}, \bibinfo {author} {\bibfnamefont {X.-Y.}\ \bibnamefont {Wang}}, \bibinfo {author} {\bibfnamefont {Z.-Y.}\ \bibnamefont {Dai}}, \bibinfo {author} {\bibfnamefont {K.}~\bibnamefont {Han}}, \bibinfo {author} {\bibfnamefont {Y.-T.}\ \bibnamefont {Wang}},\ and\ \bibinfo {author} {\bibfnamefont {T.-L.}\ \bibnamefont {Xia}},\ }\bibfield  {title} {\bibinfo {title} {{Anomalous Hall effect and topological Hall effect in the noncollinear antiferromagnet \ce{V_{0.3}NbS2}}},\ }\href@noop {} {\bibfield  {journal} {\bibinfo  {journal} {Physical Review B}\ }\textbf {\bibinfo {volume} {107}},\ \bibinfo {pages} {134436} (\bibinfo {year} {2023})}\BibitemShut {NoStop}%
\bibitem [{\citenamefont {Ueda}\ \emph {et~al.}(2012)\citenamefont {Ueda}, \citenamefont {Iguchi}, \citenamefont {Suzuki}, \citenamefont {Ishiwata}, \citenamefont {Taguchi},\ and\ \citenamefont {Tokura}}]{ueda2012topological}%
  \BibitemOpen
  \bibfield  {author} {\bibinfo {author} {\bibfnamefont {K.}~\bibnamefont {Ueda}}, \bibinfo {author} {\bibfnamefont {S.}~\bibnamefont {Iguchi}}, \bibinfo {author} {\bibfnamefont {T.}~\bibnamefont {Suzuki}}, \bibinfo {author} {\bibfnamefont {S.}~\bibnamefont {Ishiwata}}, \bibinfo {author} {\bibfnamefont {Y.}~\bibnamefont {Taguchi}},\ and\ \bibinfo {author} {\bibfnamefont {Y.}~\bibnamefont {Tokura}},\ }\bibfield  {title} {\bibinfo {title} {{Topological Hall effect in pyrochlore lattice with varying density of spin chirality}},\ }\href@noop {} {\bibfield  {journal} {\bibinfo  {journal} {Physical review letters}\ }\textbf {\bibinfo {volume} {108}},\ \bibinfo {pages} {156601} (\bibinfo {year} {2012})}\BibitemShut {NoStop}%
\bibitem [{\citenamefont {Rout}\ \emph {et~al.}(2019)\citenamefont {Rout}, \citenamefont {Madduri}, \citenamefont {Manna},\ and\ \citenamefont {Nayak}}]{rout2019field}%
  \BibitemOpen
  \bibfield  {author} {\bibinfo {author} {\bibfnamefont {P.~K.}\ \bibnamefont {Rout}}, \bibinfo {author} {\bibfnamefont {P.~P.}\ \bibnamefont {Madduri}}, \bibinfo {author} {\bibfnamefont {S.~K.}\ \bibnamefont {Manna}},\ and\ \bibinfo {author} {\bibfnamefont {A.~K.}\ \bibnamefont {Nayak}},\ }\bibfield  {title} {\bibinfo {title} {{Field-induced topological Hall effect in the noncoplanar triangular antiferromagnetic geometry of \ce{Mn3Sn}}},\ }\href@noop {} {\bibfield  {journal} {\bibinfo  {journal} {Physical Review B}\ }\textbf {\bibinfo {volume} {99}},\ \bibinfo {pages} {094430} (\bibinfo {year} {2019})}\BibitemShut {NoStop}%
\bibitem [{\citenamefont {Shiomi}\ \emph {et~al.}(2012)\citenamefont {Shiomi}, \citenamefont {Mochizuki}, \citenamefont {Kaneko},\ and\ \citenamefont {Tokura}}]{shiomi2012hall}%
  \BibitemOpen
  \bibfield  {author} {\bibinfo {author} {\bibfnamefont {Y.}~\bibnamefont {Shiomi}}, \bibinfo {author} {\bibfnamefont {M.}~\bibnamefont {Mochizuki}}, \bibinfo {author} {\bibfnamefont {Y.}~\bibnamefont {Kaneko}},\ and\ \bibinfo {author} {\bibfnamefont {Y.}~\bibnamefont {Tokura}},\ }\bibfield  {title} {\bibinfo {title} {Hall effect of spin-chirality origin in a triangular-lattice helimagnet \ce{Fe_{1.3}Sb}},\ }\href@noop {} {\bibfield  {journal} {\bibinfo  {journal} {Physical review letters}\ }\textbf {\bibinfo {volume} {108}},\ \bibinfo {pages} {056601} (\bibinfo {year} {2012})}\BibitemShut {NoStop}%
\bibitem [{\citenamefont {Liu}\ \emph {et~al.}(2017)\citenamefont {Liu}, \citenamefont {Zhang}, \citenamefont {Liu}, \citenamefont {Ding}, \citenamefont {Liu}, \citenamefont {Jafri}, \citenamefont {Hou}, \citenamefont {Wang}, \citenamefont {Ma},\ and\ \citenamefont {Wu}}]{liu2017transition}%
  \BibitemOpen
  \bibfield  {author} {\bibinfo {author} {\bibfnamefont {Z.}~\bibnamefont {Liu}}, \bibinfo {author} {\bibfnamefont {Y.}~\bibnamefont {Zhang}}, \bibinfo {author} {\bibfnamefont {G.}~\bibnamefont {Liu}}, \bibinfo {author} {\bibfnamefont {B.}~\bibnamefont {Ding}}, \bibinfo {author} {\bibfnamefont {E.}~\bibnamefont {Liu}}, \bibinfo {author} {\bibfnamefont {H.~M.}\ \bibnamefont {Jafri}}, \bibinfo {author} {\bibfnamefont {Z.}~\bibnamefont {Hou}}, \bibinfo {author} {\bibfnamefont {W.}~\bibnamefont {Wang}}, \bibinfo {author} {\bibfnamefont {X.}~\bibnamefont {Ma}},\ and\ \bibinfo {author} {\bibfnamefont {G.}~\bibnamefont {Wu}},\ }\bibfield  {title} {\bibinfo {title} {{Transition from anomalous Hall effect to topological Hall effect in hexagonal non-collinear magnet \ce{Mn3Ga}}},\ }\href@noop {} {\bibfield  {journal} {\bibinfo  {journal} {Scientific Reports}\ }\textbf {\bibinfo {volume} {7}},\ \bibinfo {pages} {515} (\bibinfo {year} {2017})}\BibitemShut {NoStop}%
\bibitem [{\citenamefont {S{\"u}rgers}\ \emph {et~al.}(2014)\citenamefont {S{\"u}rgers}, \citenamefont {Fischer}, \citenamefont {Winkel},\ and\ \citenamefont {L{\"o}hneysen}}]{surgers2014large}%
  \BibitemOpen
  \bibfield  {author} {\bibinfo {author} {\bibfnamefont {C.}~\bibnamefont {S{\"u}rgers}}, \bibinfo {author} {\bibfnamefont {G.}~\bibnamefont {Fischer}}, \bibinfo {author} {\bibfnamefont {P.}~\bibnamefont {Winkel}},\ and\ \bibinfo {author} {\bibfnamefont {H.~V.}\ \bibnamefont {L{\"o}hneysen}},\ }\bibfield  {title} {\bibinfo {title} {{Large topological Hall effect in the non-collinear phase of an antiferromagnet}},\ }\href@noop {} {\bibfield  {journal} {\bibinfo  {journal} {Nature communications}\ }\textbf {\bibinfo {volume} {5}},\ \bibinfo {pages} {3400} (\bibinfo {year} {2014})}\BibitemShut {NoStop}%
\bibitem [{\citenamefont {Kimbell}\ \emph {et~al.}(2022)\citenamefont {Kimbell}, \citenamefont {Kim}, \citenamefont {Wu}, \citenamefont {Cuoco},\ and\ \citenamefont {Robinson}}]{kimbell2022challenges}%
  \BibitemOpen
  \bibfield  {author} {\bibinfo {author} {\bibfnamefont {G.}~\bibnamefont {Kimbell}}, \bibinfo {author} {\bibfnamefont {C.}~\bibnamefont {Kim}}, \bibinfo {author} {\bibfnamefont {W.}~\bibnamefont {Wu}}, \bibinfo {author} {\bibfnamefont {M.}~\bibnamefont {Cuoco}},\ and\ \bibinfo {author} {\bibfnamefont {J.~W.}\ \bibnamefont {Robinson}},\ }\bibfield  {title} {\bibinfo {title} {{Challenges in identifying chiral spin textures via the topological Hall effect}},\ }\href@noop {} {\bibfield  {journal} {\bibinfo  {journal} {Communications Materials}\ }\textbf {\bibinfo {volume} {3}},\ \bibinfo {pages} {19} (\bibinfo {year} {2022})}\BibitemShut {NoStop}%
\bibitem [{Dat()}]{Data}%
  \BibitemOpen
  \href@noop {} {\bibinfo {title} {Data supporting the findings of the study "complex magnetic phase diagrams in \ce{Tb2IrAl4Ge2} and \ce{Er2IrAl4Ge2}"}},\ \bibinfo {howpublished} {\url{https://doi.org/10.5281/zenodo.22210799}}\BibitemShut {NoStop}%
\end{thebibliography}
\end{document}